\documentclass{jfm}
\usepackage{hyperref}
\usepackage{microtype,amsmath}

\usepackage{graphicx}
\usepackage{natbib}
\usepackage[svgnames]{xcolor}
\hypersetup{colorlinks,breaklinks,
            urlcolor=Teal,
            linkcolor=MediumBlue,
            citecolor=NavyBlue,
            pdfauthor={Shruti Mishra, Shmuel M. Rubinstein, and Chris H. Rycroft},
            pdftitle={Computing the viscous effect in early-time drop impact dynamics}
}

\newcommand{\RomanNumeralCaps}[1]
\linenumbers

\newcommand{\St}{\textit{St}}
\newcommand{\We}{\textit{We}}
\newcommand{\Sc}[1]{$S_{V=#1\text{\,m}/\text{s}}$}
\newcommand{\taudat}{\tau_\text{dat}}
\newcommand{\tzdat}{t_0^\text{dat}}

\usepackage{textcomp,nicefrac}

\definecolor{purp}{rgb}{0.4,0.2,0.8}

\newcommand{\mmsqs}{\,\text{mm}^2\!/\text{s}}
\newcommand{\vtwoptfive}{2.5\mmsqs}
\newcommand{\vfour}{4\mmsqs}
\newcommand{\vsixptfive}{6.5\mmsqs}
\newcommand{\vten}{10\mmsqs}
\newcommand{\vthirteen}{13\mmsqs}
\newcommand{\vtwenty}{20\mmsqs}
\newcommand{\vtwentyfive}{25\mmsqs}
\newcommand{\vthirtytwo}{32\mmsqs}
\newcommand{\vforty}{40\mmsqs}
\newcommand{\vhun}{100\mmsqs}
\newcommand{\vhunsixty}{160\mmsqs}
\newcommand{\vthreehun}{300\mmsqs}
\title{Computing the viscous effect in early-time drop impact dynamics}

\author{Shruti Mishra\aff{1},
  Shmuel M. Rubinstein\aff{1,2}
 \and Chris H. Rycroft\aff{1,3}\corresp{\email{chr@seas.harvard.edu}}}

\affiliation{\aff{1}John A.~Paulson School of Engineering and Applied Sciences, Harvard University, Cambridge, MA 02138, USA
\aff{2}The Racah Institute of Physics, The Hebrew University of Jerusalem, Jerusalem 91904, Israel
\aff{3}Computational Research Division, Lawrence Berkeley Laboratory, Berkeley, CA 94720, USA}

\begin{document}
\renewcommand{\sectionautorefname}{Section}
\renewcommand{\subsectionautorefname}{Section}
\renewcommand{\subsubsectionautorefname}{Section}
\maketitle

\begin{abstract}
The impact of a liquid drop on a solid surface involves many intertwined physical effects, and is influenced by drop velocity, surface tension, ambient pressure and liquid viscosity, among others. Experiments by \citet{kolinski2014lift} show that the liquid--air interface begins to deviate away from the solid surface even before contact. They found that the lift-off of the interface starts at a critical time that scales with the square root of the kinematic viscosity of the liquid. To understand this, we study the approach of a liquid drop towards a solid surface in the presence of an intervening gas layer. We take a numerical approach to solve the Navier--Stokes equations for the liquid, coupled to the compressible lubrication equations for the gas, in two dimensions. With this approach, we recover the experimentally captured early time effect of liquid viscosity on the drop impact, but our results show that lift-off time and liquid kinematic viscosity have a more complex dependence than the square root scaling relationship. We also predict the effect of interfacial tension at the liquid--gas interface on the drop impact, showing that it mediates the lift-off behavior.
\end{abstract}


\section{Introduction}
\label{sec:drop-intro}
The dynamics of a falling liquid drop as it impacts on a substrate depend on the initial conditions and physical properties of the liquid, the surrounding media and the substrate. Depending on the relevant physical parameters, the liquid drop may splash, spread or bounce upon impact with a solid substrate. The diversity of patterns produced by liquid drops impacting on a solid substrate in the presence of ambient air were first captured by \citet{worthington1877xxviii}, even before the invention of flash photography, through observations of patterns under the light of a spark produced by a break in an electric circuit when the drop hits the substrate. Worthington's drawings of these patterns sparked many investigations into the physical phenomena associated with drop impact, many of which are summarized in several review articles \citep{rein1993phenomena,yarin2006drop,josserand2016drop}.

\begin{figure}
    \centering
    \includegraphics[scale=0.5]{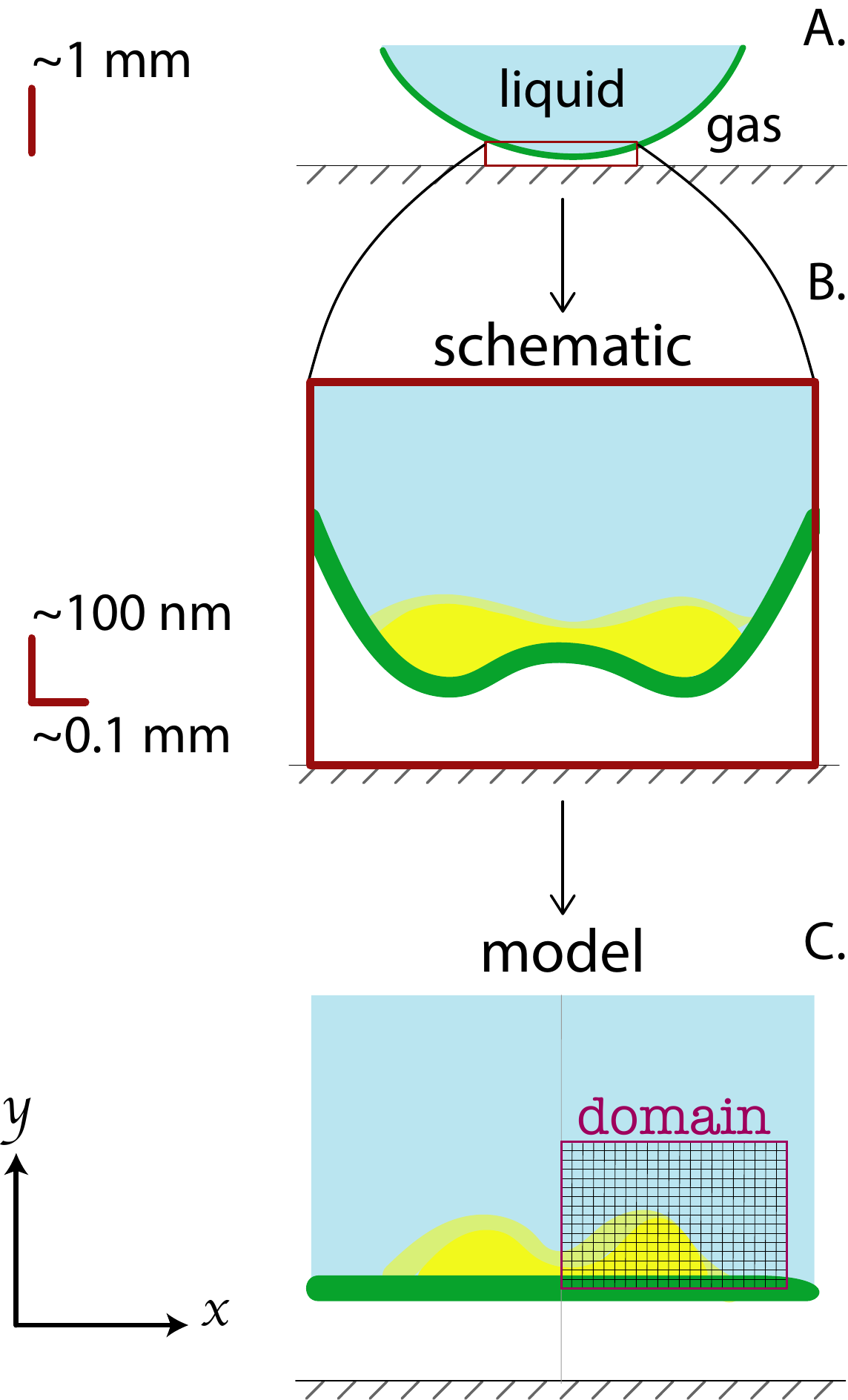}
    \caption{Schematic of the physical and computational model. (a) Relative locations of the falling drop, intervening gas layer and solid substrate. (b) A schematic zoomed into the region of interest, showing an exaggerated view of a deformed liquid--gas interface (green), and a schematic of the region where viscosity is important. (c) Control volume and grid for discretization constructed in the fluid domain.}
    \label{fig:drop-schematic}
\end{figure}

Physical parameters that are known to affect the dynamics of a drop upon impact with a substrate include the initial size and velocity of the drop \citep{pasandideh1996capillary}, viscosity of the liquid \citep{pasandideh1996capillary}, pressure and viscosity of the surrounding medium \citep{xu2005drop}, interfacial tension between the liquid and its surrounding medium \citep{rioboo2003experimental,pasandideh1996capillary}, and the physical properties of the substrate \citep{range1998influence,rein1993phenomena,yarin2006drop,josserand2016drop}. Experiments by \citet{xu2005drop} show that in the impact of a liquid drop on a solid substrate, ambient air pressure determines whether the drop eventually splashes, characterized by the ejection of a number of small drops into the air, or forms a thin film and spreads smoothly onto the surface. This suggests a role of the surrounding gaseous medium in determining the impact dynamics even before the drop makes contact with the solid substrate, motivating investigation into the role of an intervening gaseous layer in determining dynamics of a liquid drop as it approaches a solid substrate, prior to contact of the drop with the substrate. The series of theoretical investigations by Mandre, Mani and Brenner \citep{mandre2009precursors, mani2010events, mandre2012mechanism} considers an inviscid liquid drop falling towards a solid surface in the presence of an ideal gas, and predicts that the drop deforms due to a buildup of pressure in the gas trapped underneath the drop, prior to its contact with the solid substrate. Since these studies assume the fluid flow in the drop is inviscid, the velocity field is given by a potential flow. This substantially simplifies the analysis, allowing the dynamics to modeled using values of relevant field variables exclusively at the liquid--gas interface, without explicitly computing the fluid flow in the bulk of the drop. A schematic of the deformation of the drop is shown in Figures \ref{fig:drop-schematic}(a) and \ref{fig:drop-schematic}(b).

Experimental studies confirm the theoretical predictions of drops deforming on a layer of air between the liquid drop and solid substrate, prior to contact between the drop and the substrate, and show measurements of the air layer underneath the impacting drop \citep{kolinski2012skating,van2012direct,de2012dynamics}. However, subsequent experimental work by \citet{kolinski2014lift} shows the effect of liquid viscosity on the evolution of the liquid--air interface even before contact with the substrate, which is not captured by the potential flow description of the evolution of liquid dynamics in previous theoretical work \citep{mandre2009precursors,mani2010events,mandre2012mechanism}.

In this work, we are concerned with the dynamics of a liquid drop as it approaches a rigid, non-porous, solid surface, in the presence of an ambient gas. We modify the physical model of Mandre, Mani and Brenner by incorporating the viscosity of the liquid. Due to the additional viscous term in the governing equations for the liquid, the coupled interaction of the liquid and gas becomes theoretically intractable, and it is necessary to solve numerically for the fluid flow in the bulk of drop. Previous computational work on drop impact has been concerned with the dynamics of spreading \citep{eggers10} and splashing \citep{josserand03,boelens2018simulations}, whereas we focus on the dynamics of the drop in its approach towards a solid surface, prior to contact. 

We solve the incompressible Navier--Stokes equations in the drop, using a modern implementation of Chorin's projection method~\citep{chorin67,chorin68} that incorporates improvements from Almgren, Bell, Collela and coworkers~\citep{bell89,colella90,almgren96}. Our numerical model allows us to resolve the flow field and pressure in the drop, and examine the effects of many different physical parameters in ways (e.g.~viscosity, surface tension) that would be difficult to do in experiment. We validate our model using theoretical results \citep{mandre2009precursors,mani2010events,mandre2012mechanism}. Using our model, we are able to recapitulate the square-root scaling with liquid viscosity of the lift-off time observed by \citet{kolinski2014lift}. However, our results show that the precise relationship between lift-off time and liquid viscosity is more complicated. We explore the dependence of lift-off time on surface tension, initial drop velocity, and drop radius.

In the following sections, we describe the physical model and parameter regime, followed by approximations made in the simulation domain. We then describe the results from our simulations, along with comparisons with previous theoretical and experimental studies. Our numerical results are consistent with theoretical calculations \citep{mandre2009precursors} as well as experimental results \citep{kolinski2014lift}. Our results predict the effect of viscosity and interfacial tension on the dynamics of drop impact.

%~~~~~~~~~~%~~~~~~~~~~%~~~~~~~~~~%~~~~~~~~~~%~~~~~~~~~~%
%                        METHOD                        %
%~~~~~~~~~~%~~~~~~~~~~%~~~~~~~~~~%~~~~~~~~~~%~~~~~~~~~~%
\section{Model}
\label{sec:drop-model}
In this section, we first explain the physical set-up and specify the parameter regime considered in the rest of this work. Based on the physical set-up and parameter regime, we then describe the mathematical model. The mathematical model largely derives from previous theoretical work~\citep{mandre2009precursors,mani2010events,mandre2012mechanism}. We specify the predictions made by this model, experimentally-observed deviations from these predictions, and then the mathematical model used in our work. We then use this mathematical model to derive appropriate boundary conditions for the flow in the drop.

\subsection{Physical problem and parameter regime}
\label{sec:drop-physical-problem}
The physical set-up is a drop of liquid falling towards a flat, solid surface in the presence of a surrounding gas. As the liquid drop falls towards the solid surface, it interacts with the surrounding gas. We are interested in modeling and simulating the coupled dynamics of the liquid and the gas before the liquid makes contact with the solid surface.

\autoref{fig:drop-schematic}(a) shows a schematic of the physical problem, indicating the relative locations of the liquid drop, the surrounding gas and the solid surface. The drop is initially spherical, with radius $R$, falling towards a horizontal solid surface at uniform vertical velocity $V_0$. The gas is initially at uniform pressure $P_0$. We specify the values of initial conditions and physical properties considered for the liquid and air in \autoref{tbl:drop-params}. We use the subscripts $l$ and $g$ to denote the liquid drop and the gas, respectively, and $0$ to denote the initial conditions.

\begin{table}
\centering
\begin{tabular}{llll} 
 Symbol & Quantity & Order of magnitude & Units\\ \hline
 $P_0$ & Initial gas pressure & $\mathcal{O}(10^{5})$ & $\mathrm{Pa}$   \\
 $R$ & Initial drop radius & $\mathcal{O}(10^{-3})$ & $\mathrm{m}$   \\
 $V_0$ & Initial drop velocity & $\mathcal{O}(10^{-1}\text{--}10^{0})$ & $\mathrm{m/s}$   \\ \hline
 $\gamma$ & Ratio of specific heat capacities, gas & $\mathcal{O}(1)$ & --   \\
 $\rho_g$ & Density of gas & $\mathcal{O}(10^0)$ & $\mathrm{kg/m^3}$   \\
 $\rho_l$ & Density of liquid & $\mathcal{O}(10^3)$ & $\mathrm{kg/m^3}$   \\
 $\nu_g$ & Kinematic viscosity of gas & $\mathcal{O}(10^{-5})$ & $\mathrm{m^2/s}$   \\
 $\nu_l$ & Kinematic viscosity of liquid & $\mathcal{O}(10^{-6}\text{--}10^{-4})$ & $\mathrm{m^2/s}$ 
\end{tabular}
\caption{Relevant initial conditions and physical parameters, and their approximate values, which informs the mathematical model. The subscripts $l$ and $g$ denote the liquid drop and the gas, respectively, and $0$ denotes initial conditions.}
\label{tbl:drop-params}
\end{table}

A theoretical consideration of our physical set-up for the parameter regime specified in \autoref{tbl:drop-params}. In this parameter regime, the potential flow theory argues that the relevant Reynolds number, based on the initial velocity $V_0$ and drop radius $R$, is $\mathcal{O}\left(10^{2}\right)$, allowing for an inviscid consideration of the liquid. We refer to the mathematical model as the \emph{potential flow model}, and the theoretical predictions arising from this mathematical model as the \emph{potential flow theory}.

The potential flow theory predicts that, as the drop falls towards the solid surface, a build up of air pressure underneath the drop causes the drop to deform in the middle, developing a \emph{dimple}. The drop then spreads over a thin layer of gas that separates it from the substrate. The potential flow theory predicts scaling laws for the height of the gas layer. Subsequent experimental observations of the gas underneath the drop \citep{kolinski2012skating,van2012direct,de2012dynamics} show the existence of this dimple and the shape of the drop profile, similar to the predictions of the potential flow theory.

Experimental observations made by \citet{kolinski2014lift} in part of this parameter regime show that the falling drop first develops a dimple in the middle of the drop, as schematized by the liquid--gas interface shown in green in \autoref{fig:drop-schematic}(b). This is consistent with the potential flow theory \citep{mandre2009precursors,mani2010events,mandre2012mechanism}. Subsequently in time, and prior to contact of the liquid--gas interface with the solid substrate, the shape of the interface depends on the kinematic viscosity of the liquid. Specifically, there is a critical time $\tau_c$ at which the minimum height of the drop from the substrate stops decreasing, and the \emph{leading edge} of the drop begins to move away from the surface. The time $\tau$ is measured relative to the time when the drop would have made contact with the substrate if it were not deforming due to interactions with an intervening gas. \citet{kolinski2014lift} observe that $\tau_c \propto {\nu_l}^{1/2}$.

As a consequence of the discrepancy between the observations of \citet{kolinski2014lift}, who definitively show the effect of liquid viscosity on the shape of the liquid--gas interface well before contact between the liquid and the substrate, and the potential flow theory, we are specifically interested in capturing the role of liquid viscosity, on these coupled dynamics.  Based on the observations of \citet{kolinski2014lift}, we modify the potential flow model to consider a liquid described by the full Navier--Stokes equations, rather than an inviscid approximation.

\subsection{Mathematical model}
\label{sec:drop-math-model}
The dynamics of an initially spherical drop are naturally described in terms of three-dimensional spherical polar co-ordinates, and those of the gas layer in terms of three-dimensional cylindrical polar co-ordinates. In this co-ordinate system, from the potential flow theory \citep{mandre2009precursors,mani2010events,mandre2012mechanism} and the experimental observations of air layer profiles under an impacting drop \citep{kolinski2012skating,de2012dynamics,van2012direct,kolinski2014lift,kolinski2014drops,kolinski2019surfing}, we use the simplifying approximation of no azimuthal variation in the flow field of the drop or the gas.

As in the potential flow model, we further simplify our consideration by approximating the drop as initially being an infinite cylinder, with the long axis perpendicular to the substrate, rather than spherical. With these assumptions and approximations, we can describe the dynamics of the liquid and gas using two-dimensional Cartesian co-ordinates $(x,y)$. The initial height $h(x,t=T_0)$ of the liquid--gas interface is described by a parabolic profile as
\begin{equation}
  h(x,t=T_0) = H_0 + \frac{x^2}{2R},
\end{equation}
where $H_0$ is the initial height of the centre of the interface, i.e.\ at $x=0$.

We model the flow in the liquid using the incompressible Navier--Stokes equations. The continuity equation for the liquid is
\begin{equation}
\label{eqn:drop-cont}
\nabla \cdot \mathbf{u}_l = 0,
\end{equation}
and the momentum equation for the liquid is
\begin{equation}
\label{eqn:drop-mom}
    \mathbf{u}_{l,t} + \mathbf{u}_l \cdot \mathbf{\nabla}\mathbf{u}_l =
    -\frac{\mathbf{\nabla}p_l}{\rho_l} + \nu_l \nabla^2 \mathbf{u}_l,
\end{equation}
where $\mathbf{u}_l = \left(u_x, u_y\right)$, $p_l$ and $\rho_l$ are the velocity field and pressure field in the liquid, respectively.

Based on the potential flow theory \citep{mandre2009precursors,mani2010events,mandre2012mechanism} and subsequent experimental observations \citep{de2012dynamics,van2012direct,kolinski2014lift}, the gas layer is slender, with the horizontal length scale $L_x$ being much greater than the vertical length scale $L_y$, allowing for an approximation of one-dimensional flow in the gas. The height $h=h(x,t)$ of the gas layer is thus described by the compressible lubrication equation,
\begin{equation}
\label{eqn:drop-lubr}
    12 \mu_g \left(\rho_g h\right)_t = \left(\rho_g h^3 p_{g,x}\right)_x,
\end{equation}
and the equation of state is described by an adiabatic assumption,
\begin{equation}
\label{eqn:drop-adia}
    \frac{p_g}{P_0} = \left(\frac{\rho_g}{\rho_0}\right)^{\gamma}.
\end{equation}
The coupling of flows in the liquid and gas at their interface is done using Gibbs' condition for pressure,
\begin{equation}
\label{eqn:drop-gibbs}
    p_l = p_g + \sigma h_{xx},
\end{equation}
and equality of shear stress in the liquid and gas at the liquid--gas interface is
\begin{equation}
\label{eqn:drop-shear}
    \tau_l(x,h) = \tau_g(x,h).
\end{equation}
Far enough away from the deforming effects of the intervening gas layer on the liquid drop, the pressure in the gas is the ambient pressure, $\lim_{x\rightarrow \pm \infty} p_g(x,t) = P_0$.

Given that the initial and boundary conditions are symmetric about $x=0$, we modify our mathematical model to incorporate symmetry boundary conditions about $x=0$. We do so by applying the boundary conditions $\left(u,v_x\right)=(0,0)$ in the liquid domain and $p_{g,x}=0$ and $h_x=0$ in the gas layer.

\subsection{Liquid domain and boundary conditions}
Having specified the physical problem in \autoref{sec:drop-physical-problem} and the mathematical model, with initial and boundary conditions, in \autoref{sec:drop-math-model}, we now turn to a description of the domain and boundary conditions that specify the flow in the liquid in a computationally tractable manner. The assumptions that lead to numerical approximations in this section are specified in sections \ref{sec:drop-physical-problem} and \ref{sec:drop-math-model}. We detail them here as appropriate.

\subsubsection{Rectangular domain with mass flux}
Following from the previous section, the liquid drop is described by a two-dimensional flow field. As the drop interacts with the gas layer, its shape changes.
The drop begins to deform at the interface, close the the middle, at $x=0$. With this observation, we choose a region of interest, as shown in \autoref{fig:drop-schematic}(b), that stretches outwards from $x=0$ and upwards from $y=h(x,t)$.

We make an additional approximation that the flow in the liquid at $y=h(x,t)$ can be approximated by the flow at $y=0$. We can remove the spatial dependence because the liquid--gas interface is slender. We simulate the control volume with and without a temporal dependence $y=h(t)$, and do not observe a significant change in results. With these approximations, we are able to choose a rectangular, inertial control volume, stretching outwards from $x=0$ and upwards from $y=0$, whose size is chosen based on the relevant physical scales (\autoref{sec:drop-domain}). With the use of symmetry in the domain, as described in \autoref{sec:drop-math-model}, we need to model only half of the control volume. We choose to model the right half, leading to the computational domain shown in \autoref{fig:drop-schematic}(c). In subsequent sections, we describe the boundary conditions at each of the boundaries in the computational domain.

\subsubsection{Treatment of the viscous term and bottom boundary condition}
Initially, the drop is moving at uniform vertical velocity, meaning that all spatial velocity gradients are zero, and $\nabla^2 \mathbf{u}_l=0$. Deformation of the drop begins at $x=0$. Our model consequently assumes that the shear stresses are small away from this region where the drop has deformed. In a two-dimensional flow, vorticity is introduced only through shear stresses at the boundary. The vorticity therefore spreads into the interior of the drop from the deformed interface. We thus model the drop as having a region close to $(x,y)=(0,0)$, where viscosity is important, while far away from this region, $\nu_l\nabla^2\mathbf{u}_l$ makes a negligible contribution to equation \eqref{eqn:drop-mom}. A schematic of the region where viscosity is important in the deformed drop is shown by yellow in \autoref{fig:drop-schematic}(b) and (c). Consequently, we model the domain as having one boundary where viscous effects are important, which is the bottom boundary.

At the bottom boundary, we apply the boundary conditions specified in equations \eqref{eqn:drop-gibbs} and \eqref{eqn:drop-shear}. Equation \eqref{eqn:drop-shear} is enforced through a condition on the gradient of the horizontal velocity,
\begin{equation}
\label{eqn:drop-u-bottom-bc}
    u_{l,y}(x,0)=\frac{\mu_g}{\mu_l}u_{g,y}(x,0).
\end{equation}

\subsubsection{Boundaries in the interior of the drop: top and right boundary conditions}
\label{sec:drop-bc-interior-analytical}
According to our approximation of a rectangular domain, the bottom boundary of the liquid interacts with the gas layer. The other boundaries are chosen to be in the interior of the drop, away from the effect of vorticity from the bottom boundary, and can be treated as inviscid. With this approximation, at these boundaries, denoted as \emph{interior boundaries}, we simplify the momentum equation for the liquid, equation \eqref{eqn:drop-mom}, by noting that the viscous term, $\nu_l\nabla^2\mathbf{u}_l$ is relatively small. 

Additionally, the velocity gradients in the liquid are initially zero. Away from the effects of the gas layer, the nonlinear advective term in the momentum equation, ${\bf u}_l \cdot \nabla {\bf u}_l$ in equation \eqref{eqn:drop-mom}, will also be small. With these approximations, we obtain a simplified momentum equation for the liquid at these boundaries,
\begin{equation}
\label{eqn:drop-mom-simp}
    \mathbf{u}_{l,t} =
    -\frac{\mathbf{\nabla}p_l}{\rho_l}.
\end{equation}
We use this simplified momentum equation to obtain velocity boundary conditions at the interior boundaries. Taking the divergence of equation \eqref{eqn:drop-mom-simp} and noting that the flow in the liquid is incompressible, we obtain $\nabla^2 p_l = 0$. Knowing the pressure at the bottom boundary, and approximating the liquid as a semi-infinite domain, we get an analytical expression for the pressure at the interior boundaries,
\begin{equation}
\label{eqn:drop-p-int}
    p_l(x,y) = \frac{1}{\pi}\int_{-\infty}^{\infty}p_l(s,0) \frac{y}{(x-s)^2+y^2} \ \mathrm{d}s.
\end{equation}
Assuming that the domain is larger than the region where the pressure from the gas is non-zero, we numerically integrate over a finite length of the domain boundary to get the pressure at a particular $(x,y)$ location along the interior boundaries. Given the pressure at the boundary, we integrate equation \eqref{eqn:drop-mom-simp} in time to obtain velocity boundary conditions at the interior boundaries. The size of the region where vorticity is significant determines the size of the domain, and the locations of the interior boundaries, which are calibrated in simulation.

\subsection{Choice of domain}
\label{sec:drop-domain}
We simulate the coupled dynamics of the liquid and gas over the time interval $0\le t \le t^\text{end}$, for a simulation domain of $0 \le x \le L$ and $0 \le y \le \beta L$, where $\beta$ is the aspect ratio, for the liquid, and a domain of $0 \le x \le L$ for the gas. To set the size of the domain and duration of the simulation, we consider the initial dynamics as the drop falls toward the surface, and compare to the theoretical results of the potential flow theory \citep{mandre2009precursors}. We use the scaling results of the potential flow theory for the centre of the gas layer, $h(x=0,t)$. The scaling analyses predict the height $H^*$ at which the pressure buildup in the gas layer causes the drop to undergo significant deformation and develop a stagnation point at $x=0$, forming a dimple.

According to potential flow theory, the compressibility of the flow in the gas is determined by the parameter 
\begin{equation}
  \epsilon = \frac{P_0 R \St^{4/3}}{\mu_g V},
\end{equation}
which is the ratio of the initial gas pressure to the pressure that would have built up in the gas were treated as incompressible.

The results of \citet{kolinski2014lift}, showing the relationship between $\nu_l$ and the drop profile, correspond to $\epsilon^{-1} < 1$, meaning that the flow in the gas can be considered to be incompressible. While our simulations model the full incompressible flow in the gas, we use this approximation to set the initial conditions. In this regime, we have the estimate
\begin{equation}
  \label{eq:drop-hstar}
  H^* = R \St^{2/3}
\end{equation}
where $\St=\mu_g/(\rho_l V R)$ is defined as the inverse of a modified Stokes number.\footnote{The inverse of the modified Stokes number, $\St=\mu_g/(\rho_l V R)$, has properties of two different media, where the usual Stokes number is defined for a single medium.}
This scaling result is strictly valid for inviscid flow, $\nu_l = 0$, $\Rey\rightarrow{}\infty$, and no surface tension at the liquid--air interface, $\sigma = 0$, $\We\rightarrow{}\infty$. While our simulations incorporate liquid viscosity and surface tension, equation \eqref{eq:drop-hstar} still provides a good estimate for the height at which the stagnation point forms, and is useful measure of the physical scales of interest. We therefore set the initial height to be a multiple of $H^*$, so that
\begin{equation}
  \label{eq:drop-hzero}
  H_0 = \tilde{H}_0 H^* = \tilde{H}_0 R \St^{2/3}
\end{equation}
for a dimensionless constant $\tilde{H}_0$. Similarly, we set \smash{$t_\text{end}= \tilde{t}_\text{end} R \St^{2/3}/V$}, and based on the parabolic shape of the initial profile we choose $L= \tilde{L} R \St^{1/3}$, where $\tilde{L}$ and $\tilde{t}_\text{end}$ are dimensionless constants.

%~~~~~~~~~~%~~~~~~~~~~%~~~~~~~~~~%~~~~~~~~~~%~~~~~~~~~~%
%                     COMPUTATIONS                     %
%~~~~~~~~~~%~~~~~~~~~~%~~~~~~~~~~%~~~~~~~~~~%~~~~~~~~~~%

\section{Numerical implementation}
In this section, we describe the numerical implementation of the mathematical model described in \autoref{sec:drop-math-model}. Throughout the liquid domain, we compute and record the liquid velocity $\mathbf{u}_l$ and pressure $p_l$. In the gas layer, compute and track the height $h$ and pressure $p_g$. A reader may skip the remainder of section without loss of continuity.

The flow-field variables $\phi$ are computed at discretized time steps, with simulation time $t=n \Delta t$, where $\Delta t$ is the time step discretization used in the simulation, and $n$ is an integer counter. Given values of field variables $\phi^{(n)}=\phi(t=n \Delta t)$, the goal of the simulation is to compute the field variables $\phi^{(n+1)}$. This section describes the numerical method for doing so, for the liquid flow-field variables, $\mathbf{u}_l$ and $p_l$, and gas flow-field variables, $h$ and $p_g$.

\subsection{Projection method}
\label{sec:drop-projection}
The core of the algorithm is based on Chorin's projection method \citep{chorin67,chorin68}, with further developments in the computational techniques behind the fluid simulation method \citep{bell89,almgren98}. In particular, the work of \citet{almgren96} introduces the approximate projection method discretized using the finite-element method (FEM). The numerical methods are described in detail by \citet{sussman99}, \citet{yu2003coupled,yu07}, and \citet{rycroft20}. Here, we sketch the main features of these methods.

We solve for $\mathbf{u}_l$ and pressure $p_l$ in the liquid domain by solving the momentum equation for the liquid, equation \eqref{eqn:drop-mom}, subject to incompressibility of the liquid, as specified in equation \eqref{eqn:drop-cont}. In describing the method for the solution of the field variables in the liquid domain, for the remainder of this section, we drop the subscript $l$.

\begin{figure}
    \centering
    \includegraphics[scale=0.5]{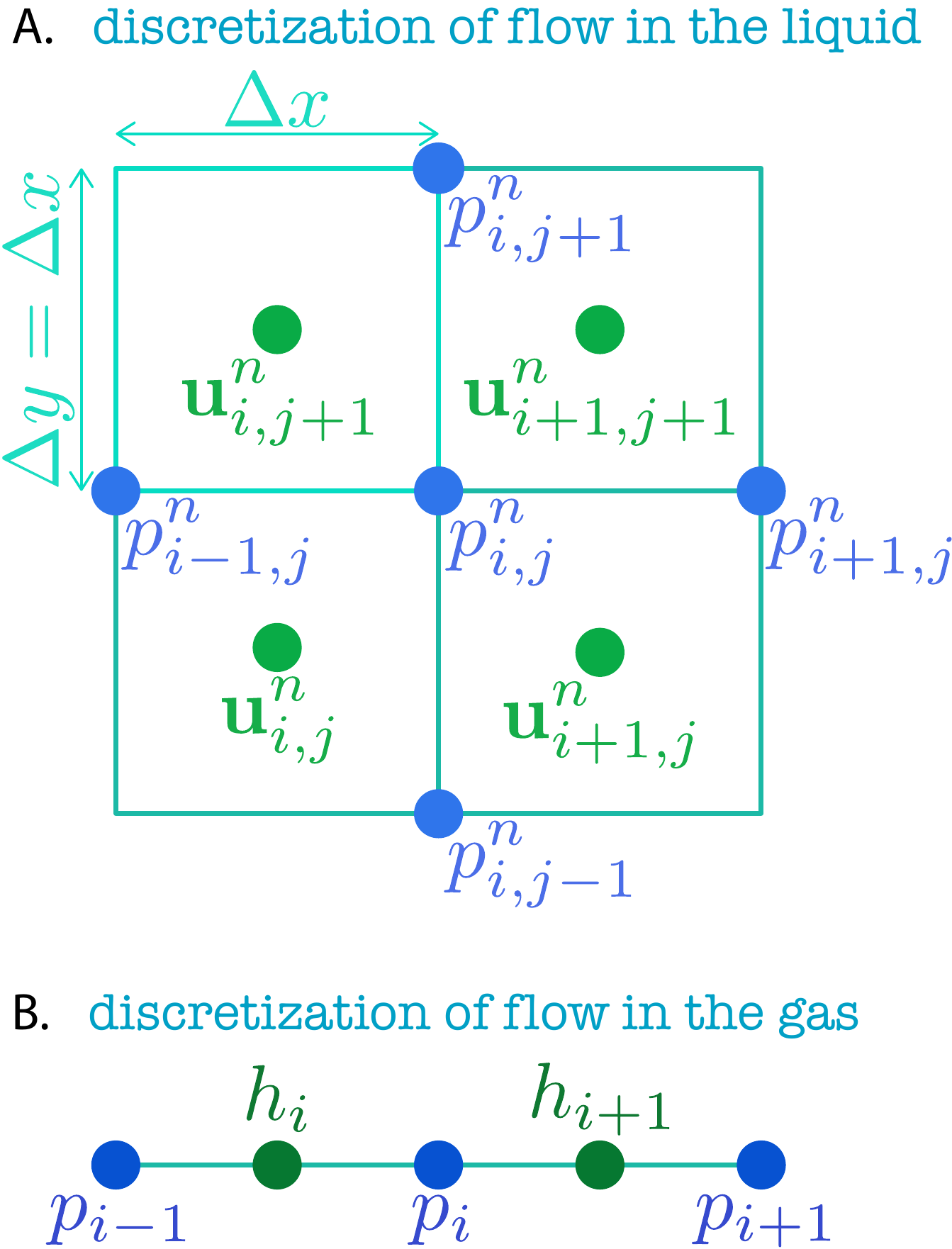}
    \caption{{\bf Discretization of the field variables.} A: The two-dimensional computational grid for the liquid. B: The one-dimensional computational grid for the gas.}
    \label{fig:drop-discretisation}
\end{figure}

The field variables are defined on a two-dimensional grid, discretized into rectangular cells of size $\Delta x$ by $\Delta y$, as shown in \autoref{fig:drop-discretisation}. The velocities $\mathbf{u}^n$ are located in the cell-centers, and the pressures $p^n$ are located at the cell corners. To advance from step $n$ to $n+1$, we first compute an intermediate velocity $\mathbf{u}^*$ according to
\begin{equation}
\label{eqn:drop-mom-discrete}
\frac{{\bf u}^{*}-{\bf u}^n}{\Delta t} = - \left[{\bf u} \cdot \nabla {\bf u}\right]^{n+1/2}  + \frac{\nu}{2} \left(\nabla^2 {\bf u}^n + \nabla^2 {\bf u}^* \right).
\end{equation}
Here, the viscous stress terms $\nu \nabla^2 \mathbf{u}$ and $\nu \nabla^2 \mathbf{u}^*$ evaluated using a standard five-point finite-difference stencil. Due to the presence of the ${\bf u}^*$ on the right hand side of equation \eqref{eqn:drop-mom-discrete} it must be solved implicitly. We use a multigrid method to solve the resulting linear system.

The advective term $[\mathbf{u} \cdot \nabla \mathbf{u}]^{n+1/2}$ is evaluated at the half-timestep $n+1/2$ using the second-order Godunov upwinding scheme of \citet{colella90}. This is performed by extrapolating the velocity in each cell to midpoints of the four edges using a first-order Taylor expansion. Doing this requires that the gradient of the velocity is first evaluated at each cell center, which is done using the fourth-order monotonicity-limited scheme of \citet{colella85}. This scheme uses a five-point stencil in each coordinate, therefore requiring information from two grid points in each orthogonal direction.

After this procedure, each edge has two velocities from the two adjacent cells. A Godunov upwinding procedure is then used to select one based on the direction of the velocity. An intermediate marker-and-cell (MAC) projection is then used to adjust the velocities to satisfy a discrete zero divergence criterion, so that the net mass flow out of each cell is zero~\citep{bell91,sussman99}. After this, the advective term can be accurately evaluated using centered finite differences of the upwinded edge-based velocities~\citep{sussman99,yu2003coupled,yu07,rycroft20}.

The velocity at step $n+1$ can be calculated from the intermediate velocity by evaluating the pressure gradient term,
\begin{equation}
\label{eqn:drop-mom-discrete2}
\frac{{\bf u}^{n+1}-{\bf u}^*}{\Delta t} = -\frac{1}{\rho}{\nabla p^{n+1}},
\end{equation}
but this requires knowledge of the pressure field $p^{n+1}$, and there is no explicit update equation for this field in the incompressible limit. To proceed, we take divergence of equation \eqref{eqn:drop-mom-discrete2} and enforce that $\nabla \cdot \mathbf{u}^{n+1}=0$, according to equation \eqref{eqn:drop-cont}. Hence,
\begin{equation}
\label{eqn:drop-pressure-poisson}
\nabla \cdot \left[\Delta t \left\{ \frac{1}{\rho} \nabla p^{n+1} \right\}\right] =  \nabla \cdot {\bf u}^*,
\end{equation}
which is a Poisson equation for the pressure that can be solved. Once $p^{n+1}$ is known, equation \eqref{eqn:drop-mom-discrete2} can be used to evaluate $\mathbf{u}^{n+1}$ and complete the timestep from $n$ to $n+1$. We solve equation \eqref{eqn:drop-pressure-poisson} using the FEM discretization of \citet{almgren96}. Here, each pressure value at a cell corner has a corresponding bilinear hat function over the four neighboring grid cells~\citep{yu2003coupled,yu07,rycroft20}. Once the pressure is computed, the gradient in equation \eqref{eqn:drop-mom-discrete2} is evaluated using centered finite differences of the pressure field values.

\subsection{Implementation of boundary conditions: ghost layers}
The boundary conditions are implemented by means of ghost layers. Around the boundaries of the two-dimensional liquid domain, there are two layers of additional grid points. Two layers are required in order to evaluate the fourth-order monotonicity-limited derivatives arising from the advective term. The boundary conditions are specified in these layers of grid points. These conditions are then used to compute the necessary gradients in equation \eqref{eqn:drop-mom-discrete}.

The velocity boundary conditions at the top and right boundaries of the liquid domain, corresponding to the interior of the drop, and described in \autoref{sec:drop-bc-interior-analytical}, are obtained through substituting the expression for pressure given by equation \eqref{eqn:drop-p-int} in the expression for velocity at these boundaries, equation \eqref{eqn:drop-mom-simp}, and then using Simpson's rule to numerically integrate the pressure along these boundaries.

\subsection{Solution of the gas layer equations}
The field variables $h$ and $p_g$ for the gas layer are discretized in accordance with the discretization for the field variables for the liquid. Specifically, the pressure field $p_g$ is stored at grid points in the same locations along the horizontal axis as the pressure field in the liquid, as shown in \autoref{fig:drop-discretisation}. The height $h$ is stored at grid points in the same locations along the horizontal axis as the velocity field $\mathbf{u}$ in the liquid.

As shown in Appendix \ref{app:drop-gas_eq_deriv}, substituting the equation of state \eqref{eqn:drop-adia} into the lubrication equation \eqref{eqn:drop-lubr} yields
\begin{equation}
  \label{eqn:drop-gas-rewrite}
12 \frac{\mu}{\gamma} h p_{t} + 12 \mu h_t  p
= \frac{1}{\gamma} h^3 p_{x}  p_{x} + 3 h^2 h_x p p_{x}  +h^3 p p_{xx},
\end{equation}
where the $g$ subscript on the pressure has been dropped. We solve for this equation as follows:
\begin{enumerate}
\item $p$ is known at the time step $t^{(n)} = n \ \Delta t $
\item $h$, $h_t$, and therefore $h_x$ are known at time step $t^{(n)} = n \ \Delta t $
\item solve for $p$ at time step $t^{(n+1)} = (n+1) \ \Delta t$
\item use $p$ as a boundary condition for the flow in the liquid, to obtain $v=h_t$ at time step $t^{(n+1)} $
\end{enumerate}
To numerically solve equation \eqref{eqn:drop-gas-rewrite} we discretize the spatial
derivatives using second-order centered finite differences. To avoid a timestep
restriction, we make use of the backward Euler method for discretizing the
temporal derivative. Hence we obtain the discretized form
\begin{multline}
  12 \frac{\mu}{\gamma} \bar{h} \left(\frac{p^{n+1}_i - p^n_i}{\Delta t} \right) + 12 \mu \bar{h}_t p^{n+1}_i
  = \frac{1}{\gamma} \bar{h}^3 \left(\frac{p^{n+1}_{i+1}-p^{n+1}_{i-1}}{2 \Delta x} \right)^2 \\
  + 3 \bar{h}^2 \bar{h}_x p^{n+1}_i \left(\frac{p^{n+1}_{i+1}-p^{n+1}_{i-1}}{2 \Delta x} \right)
  + \bar{h}^3 p^{n+1}_i \left(\frac{p^{n+1}_{i+1}-2 p^{n+1}_{i} + p^{n+1}_{i-1}}{\Delta x^2} \right),
  \label{eqn:drop-gas-fd}
\end{multline}
that can be solved for the gas pressure $p^{n+1}_i$. In equation
\eqref{eqn:drop-gas-fd}, the terms $\bar{h}$, $\bar{h}_t$, and $\bar{h}_x$ must be
evaluated at the location of $p^{n+1}_i$. Since the height field is staggered
with respect to the pressure field, this is done via linear interpolation and
centered differencing, so that
\begin{equation}
  \bar{h} = \frac{h_i+h_{i+1}}{2}, \qquad \bar{h}_t = \frac{h_{t,i} + h_{t,i+1}}{2}, \qquad \bar{h}_x = \frac{h_{i+1}-h_i}{\Delta x}.
\end{equation}
Due to the products of pressure terms on the right hand side of equation
\eqref{eqn:drop-gas-fd}, it is a nonlinear system of equations. We solve this using
the Newton--Raphson method as described in Appendix \ref{app:drop-gas_eq_numerics}.

\subsection{Code implementation and parameter choices}
\label{sec:drop-code_impl}
The simulations are performed using a custom code written in C++ using the
OpenMP library for multithreading. The code makes use of the
Templated Geometric Multigrid (TGMG) library for solving the linear systems
that arise when solving for the fluid. Each timestep
requires four linear system solves: (1) to apply the MAC projection, (2) to
apply the approximate FEM projection, and (3,4) to solve for the $x$ and $y$
velocity components when treating the viscous stress term implicitly. The code
accepts a text configuration file as input, which sets all of the physical
parameters, and describes the computational domain. The code and sample text
configuration files are available on GitHub---see the data availability statement. The simulations are
performed on a $M \times N$ grid for the liquid domain, and grid of length $M$
for the gas layer. We choose the grid spacings to be equal, so that $\Delta
x=\Delta y$. This implies that the aspect ratio $\beta$ is equal to $N/M$.

Since the second derivative terms in both the liquid domain and the gas layer
are handled implicitly, and the surface tension term in \eqref{eqn:drop-gibbs} is
small, there is no timestep restriction in the simulation that scales like
$\Delta x^2$~\citep{heath02}. We therefore choose a candidate timestep to
satisfy $\Delta t = \zeta \Delta x$ for a dimensionless constant $\zeta$, based
on satisfying Courant--Friedrichs--Lewy conditions~\citep{courant67} for the
advective terms in the liquid domain and gas layer.

The simulation outputs $N_f$ frames over the duration of the simulation. Thus
the time between frames is $t_f = t_\text{end}/N_f$. In general, an integer
multiple of candidate timesteps will not exactly match $t_f$, so that $c\Delta
t < t_f < (c+1)\Delta t$. Because of this, the actual timestep is slightly
adjusted to $\Delta t' = t_f/(c+1)$ and $c+1$ timesteps are taken between
frames. Several examples demonstrating the performance of the code are provided
in Appendix \ref{app:drop-perf}.

\begin{table}
  \centering
  \small
  \begin{tabular}{llll} 
    & Quantity & Value & Units\\ \hline
   $P_0$ & Initial gas pressure & $10^5$ & $\mathrm{Pa}$   \\
   $R$ & Initial drop radius & $1.5 \times 10^{-3}$ & $\mathrm{m}$ \\
   $V$ & Initial drop velocity & $0.45$ & $\mathrm{m/s}$ \\
   \hline
   $\gamma$ & Ratio of specific heat capacities, gas & $1.4$ & --   \\
   $\sigma$ & Interfacial surface tension & $0.072$ & $\mathrm{N/m}$   \\
   $\rho_g$ & Density, gas & $1.2754$ (density of air at 20 $^\circ$C)& $\mathrm{kg/m^3}$   \\
   $\rho_l$ & Density, liquid & $997.96$ & $\mathrm{kg/m^3}$   \\
   $\nu_g$ & Kinematic viscosity, gas & $1.506 \times 10^{-5}$ & $\mathrm{m^2/s}$   
  \end{tabular}
  \caption{Baseline choices for the physical parameters used in the simulations. These parameters
  are used throughout the paper, with modifications to them noted in the text.}
\label{tbl:drop-xu-exp-pot}
\end{table}

%~~~~~~~~~~%~~~~~~~~~~%~~~~~~~~~~%~~~~~~~~~~%~~~~~~~~~~%
%                        RESULTS                       %
%~~~~~~~~~~%~~~~~~~~~~%~~~~~~~~~~%~~~~~~~~~~%~~~~~~~~~~%
\section{Results and discussion}
\subsection{Initial dynamics and comparison with potential flow theory}
\label{sec:drop-hstar_comp}
To validate our approach, we first simulate the initial dynamics and we compare with the scaling results of potential flow theory that were introduced in \autoref{sec:drop-domain} for the height of the stagnation point $H^*$~\citep{mandre2009precursors,mani2010events,mandre2012mechanism}. We use the baseline physical parameters given in \autoref{tbl:drop-xu-exp-pot}. Since we only want to resolve initial deceleration of the drop, and not the formation of the thin gas layer, we use the computational parameters in column 3 of \autoref{tbl:drop-sim_params}, which feature a relatively coarse numerical grid of size $2048\times 256$. To compute $H^*$, we examine the sequence of height profiles from the simulation and find the time $t^*$ when $h_t(0,t^*)=0$, from which $H^* = h(0,t^*)$.

\begin{table}
  \centering
  \small
  \begin{tabular}{lllll}
    & Quantity & \shortstack{Initial dynamics, \\ all values of $\nu_l$} & \shortstack{Lift-off, \\ $\nu_l \le \vtwenty$} & \shortstack{Lift-off, \\ $\nu_l > \vtwenty$} \\ \hline
    $(M,N)$ & Grid dimensions & $(2048,256)$ & $(5120,960)$ & $(5120,960)$ \\
    $\beta$ & Aspect ratio & \nicefrac18 & \nicefrac{16}3 & \nicefrac{16}{3} \\
    $\zeta$ & Timestep multiplier & $8\times 10^{-3}$ & $8\times 10^{-3}$ & $8\times 10^{-3}$\\
    $\tilde{L}$ & Rescaled domain width & $30$ & $30$ & $45$ \\
    $\tilde{H}_0$ & Rescaled initial drop height & $15$ & $15$ & $15$ \\
    $\tilde{t}_\text{end}$ & Rescaled duration & $50$ & $50$ & $100$ \\
    $N_f$ & Number of frames & $250$ & $250$ & $250$
  \end{tabular}
  \caption{Baseline choices for the simulation parameters, which are all
  non-dimensional. The first set of parameters are for computing the initial
  dynamics and value of $H^*$ in \autoref{sec:drop-hstar_comp}. The second
  two sets are for the lift-off calculations in all other sections. The
  procedure for connecting the rescaled variables $\tilde{L}$, $\tilde{H}_0$,
  and $\tilde{t}_\text{end}$ to physical values is described in
  \autoref{sec:drop-domain}. 
  \label{tbl:drop-sim_params}}
\end{table}

As described in \autoref{sec:drop-domain}, at low initial velocities the flow in
the gas layer can be treated as incompressible, and the scaling result
\smash{$H^* = R \text{St}^{2/3}$} can be derived. \citet{mani2010events} extend this analysis to look at higher initial
velocities, where compressibility of the gas becomes important. This happens at
a height of
\begin{equation}
  \label{eq:drop-hcompr}
  H_* = R(\mu_g V/R P_0)^{1/2},
\end{equation}
where the subscript `$*$' is used to distinguish from the stagnation height. By
modeling the subsequent drop deceleration below $H_*$ assuming gas
compressibility, \citet{mani2010events} derive the result
\begin{equation}
  \label{eq:drop-hcomp}
  H^*_\text{comp} = R \St^{2/3} \epsilon^{(2-\gamma)/(2\gamma-1)}
\end{equation}
for the stagnation height.

\begin{figure}
    \centering
    \includegraphics[width=1\linewidth]{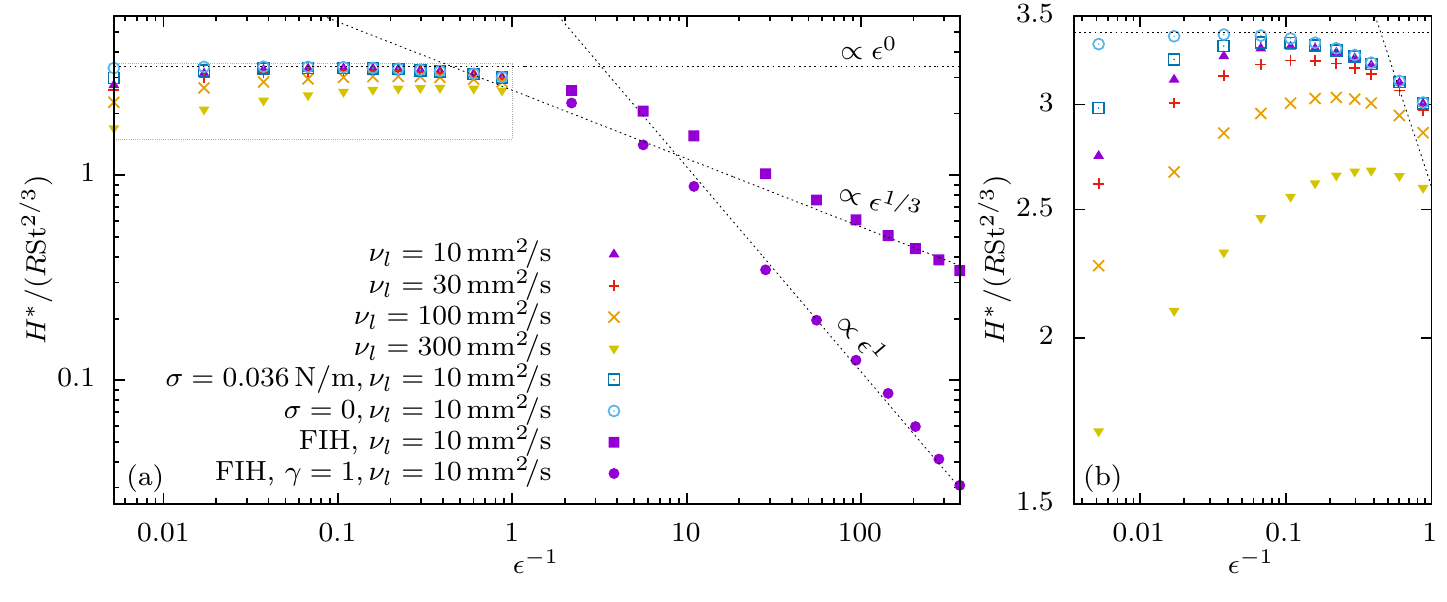}
    \caption{(a) Plot of rescaled initial stagnation height $H^*/(R\St^{2/3})$ as a function of the inverse of the dimensionless compressibility parameter $\epsilon = P_0 R \St^{4/3}/\mu_g V$, for a range of different liquid viscosities $\nu_l$, surface tensions $\sigma$, and gas parameters $\gamma$. When not otherwise stated, baseline parameters from \autoref{tbl:drop-xu-exp-pot} are used. By default, the drop starts from a height $H_0$ scaled according to equation \eqref{eq:drop-hzero}. For $\nu_l=\vten$, to account for compressibility effects, data from two simulation sequences that use a fixed initial height (FIH) based on substituting $V=2\,\text{m}/\text{s}$ into equation \eqref{eq:drop-hzero} are shown. (b) Zoomed-in plot of the same data, showing the region bounded by the dotted gray rectangle in panel (a).\label{fig:drop-hstar_scaling}}
\end{figure}

We first performed sequences of simulations using low initial velocites over
the range from $V=0.15\,\text{m}/\text{s}$ to $V=1.35\,\text{m}/\text{s}$,
using four different liquid viscosities from $\nu_l = \vten$ to $\nu_l =
\vthreehun$. We also ran two sets of simulations with the surface tension set
to half and zero its baseline value. \autoref{fig:drop-hstar_scaling}(a) shows
a plot of \smash{$H^*/(R\St^{2/3})$} as a function of $\epsilon^{-1}$, with
these six sets of data points in the left half of the domain where
$\epsilon^{-1}<1$ and gas compressibility is not important. We see that
\smash{$H^*/(R\St^{2/3})$} is approximately constant and is in agreement with
equation \eqref{eq:drop-hstar}. To examine the trends more clearly,
\autoref{fig:drop-hstar_scaling}(b) shows a zoomed-in plot of the same data. As
expected, the best agreement is achieved for the smallest value of $\nu_l$ and
for zero surface tension. At small $\epsilon^{-1}$ the values of
\smash{$H^*/(R\St^{2/3})$} are slightly lower for the cases of larger $\nu_l$
and $\sigma$. This hints at the limits of the potential flow theory
predictions~\citep{mandre2009precursors,mani2010events}.

We also performed two sequences of simulations with higher initial velocities
from $V=1\,\text{m}/\text{s}$ to $V=18\,\text{m}/\text{s}$ using
$\nu_l=\vten$, to examine the compressible regime. However, we note
from equation \eqref{eq:drop-hzero} that our usual choice for initial height scales
according to $V^{-2/3}$ whereas from equation \eqref{eq:drop-hcompr}, the height
where compressibility is important scales according to $V^{1/2}$. To fully
capture the compressible effects, the drop must start higher than $H_*$, and
thus the usual initialization procedure is problematic for large $V$.
For these simulations we therefore set $H_0$ at a fixed value based on
using $V=2\,\text{m}/\text{s}$ in equation \eqref{eq:drop-hzero}. We ran two
sequences of simulations using the usual choice of $\gamma=1.4$, and another
with $\gamma=1$; as shown in \autoref{fig:drop-hstar_scaling} these results are in
very good agreement with the scalings of $\epsilon^{1/3}$ and $\epsilon$,
respectively, that are expected from equation \eqref{eq:drop-hcomp}.

\subsection{Effect of liquid viscosity on drop profile}
\label{sec:drop-visc_liftoff}
With the initial dynamics validated, we now turn attention to simulating the continued spreading of the liquid drop on a layer of gas, and the effect of viscosity on the evolution of the liquid--gas interface. We use the same baseline physical parameters given in
\autoref{tbl:drop-xu-exp-pot}, but since we must now accurately resolve the gas
layer as it becomes very thin, we increase the simulation resolution as shown
in columns 4 \& 5 of \autoref{tbl:drop-sim_params}. Furthermore, since the deviation of the interface from the solid surface happens later for high viscosities \citep{kolinski2014lift}, we use a larger domain and longer
duration when $\nu_l > \vtwenty$, as indicated in column 5 of
\autoref{tbl:drop-sim_params}. We begin by using the baseline initial drop velocity
of $V=0.45\,\text{m}/\text{s}$ to match the experiments of \citet{kolinski2014lift}.

\begin{figure}
  \centering
  \includegraphics[width=\linewidth]{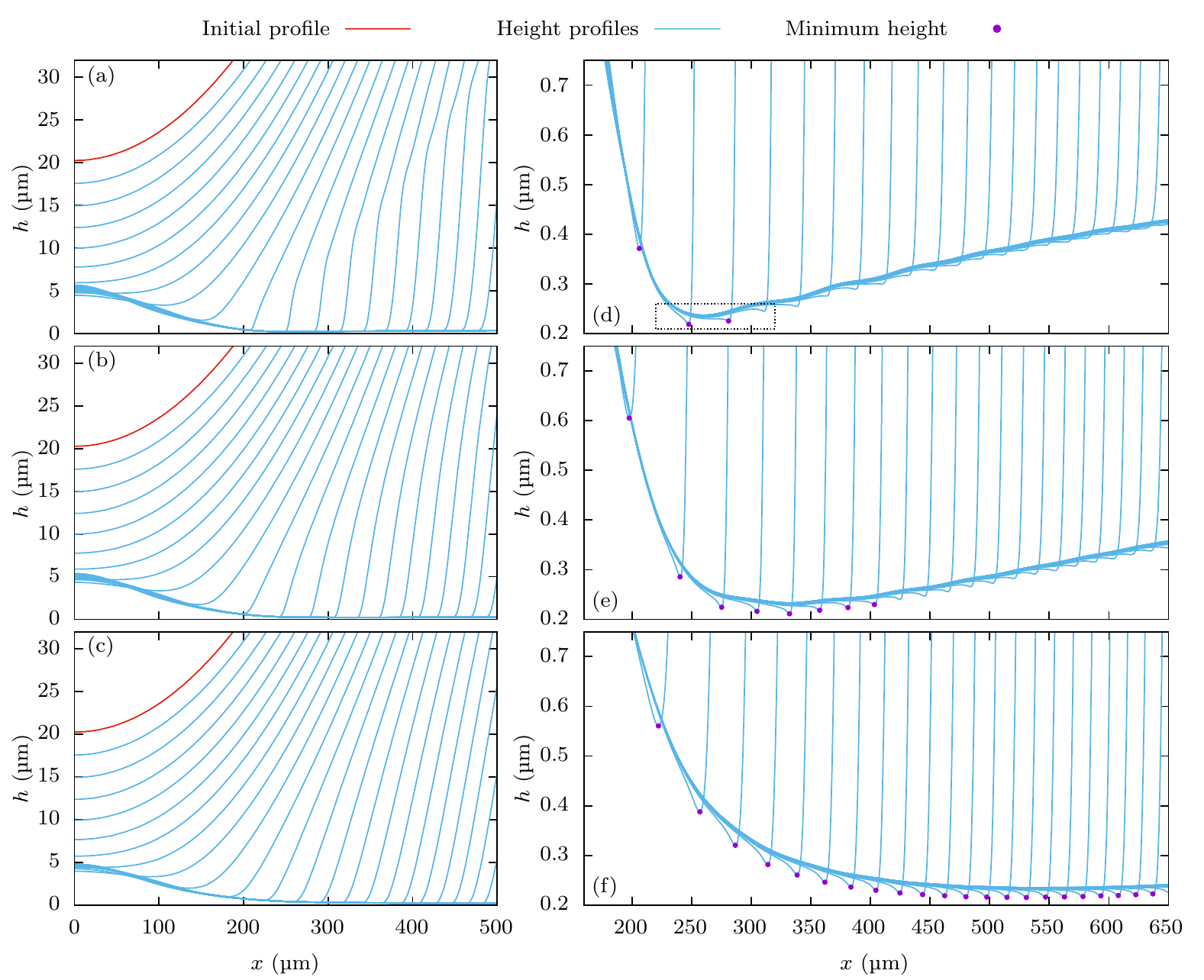}
  \caption{Profiles of the height $h(x,t)$ of the gas layer at intervals spaced $6.004\,\text{\textmu{}s}$ apart, corresponding to an integer multiple of the frame output interval $t_f$ described in \autoref{sec:drop-code_impl}, for liquid viscosities of (a) $\nu_l = \vten$, (b) $\nu_l=\vthirtytwo$, and (c) $\nu_l=\vhun$. Panels (d), (e), and (f) show the same data as (a), (b), and (c), respectively, but with a smaller range of $h$ to highlight the lift-off behavior. For each profile, the global minimum, which follows the leading tip, is also plotted on the curves; once the global minimum is no longer at the leading tip, it is no longer plotted. The dashed box in panel (d) marks a further zoomed-in region shown in \autoref{fig:drop-profiles_zoom}. \label{fig:drop-profile}}
\end{figure}

\autoref{fig:drop-profile} shows the height profiles for three different simulations
using liquid viscosities of $\nu_l = \vten$, $\nu_l = \vthirtytwo$,
and $\nu_l = \vhun$. Panels (a),
(b), and (c) show a large-scale view, where the initial parabolic profile
approaches the surface, begins to decelerate and deforms to create the dimple.
After the drop continues to fall, a thin layer of gas from $x\gtrapprox 200 \,
\text{\textmu{}m}$ is created and spreads outward. In all three simulations,
the height profiles have a front that sweeps outward as more liquid approaches
the surface.

\begin{figure}
  \centering
  \includegraphics[width=\linewidth]{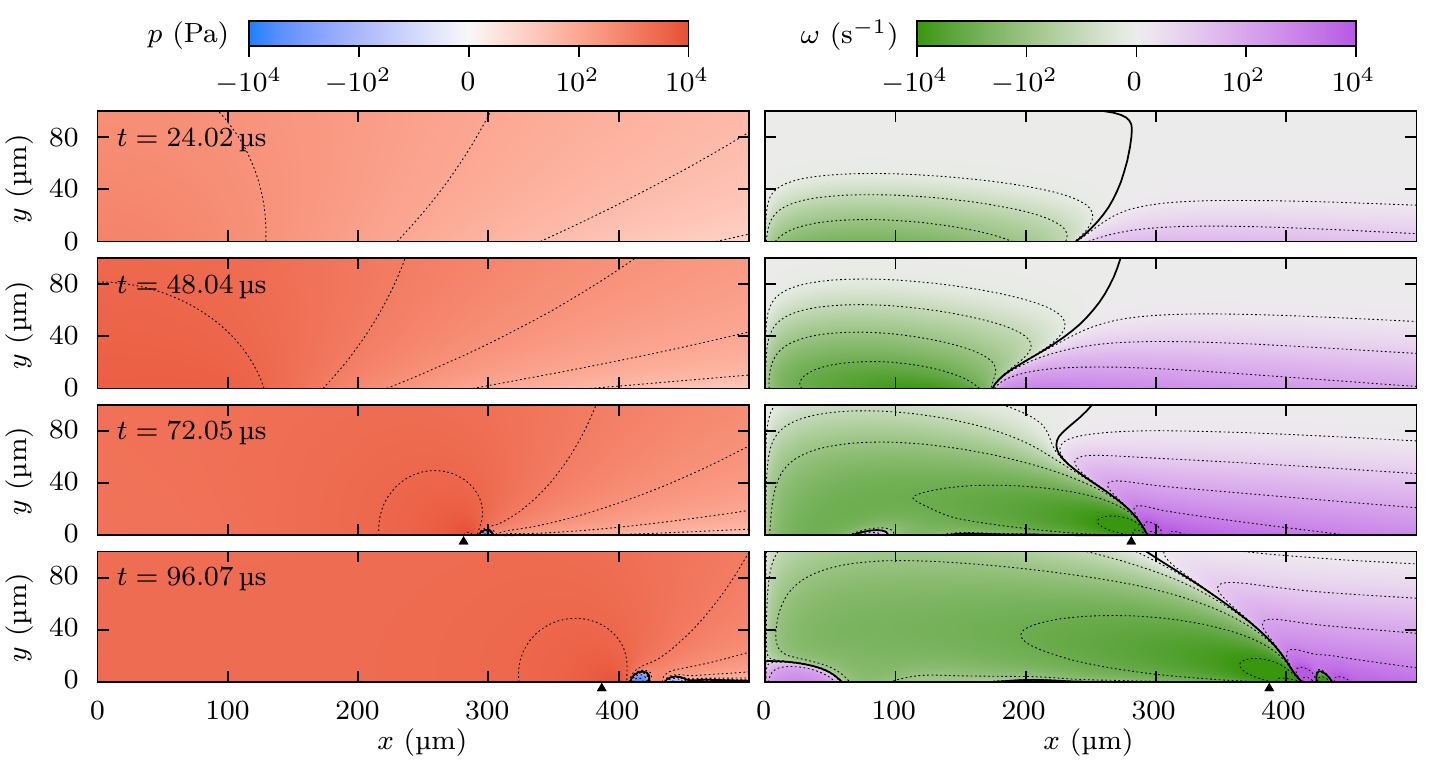}
  \caption{Snapshots of pressure $p$ (left) and vorticity $\omega=[\nabla
  \times \mathbf{u}]_3$ for a portion of the liquid domain for a simulation
  with liquid viscosity $\nu_l=\vten$. The field values exhibit large variations in
  scale and also switch sign, so a nonlinear mapping
  \smash{$f(\alpha) = \frac{1}{\log 10} \sinh^{-1} \frac{\alpha}{2}$} is used
  to create the color maps. The thick black lines indicate the zero contour.
  The thin black lines show contours for $\pm 10^{n/2}\,\text{Pa}$ for pressure
  and $\pm 10^n\,\text{s}^{-1}$ for vorticity, where $n\in \mathbb{N}_0$. For
  $t=72.05\,\text{\textmu{}s}$ and $t=96.07\,\text{\textmu{}s}$ the position of
  the front (given by the local minimum of the profiles in
  \autoref{fig:drop-profile}(a,d)) is marked with a
  triangle.\label{fig:drop-lo_vis10_pw}}
\end{figure}

\autoref{fig:drop-lo_vis10_pw} shows snapshots of the pressure and vorticity in the
liquid domain for the simulation with with $\nu_l=\vten$. At
$t=24.02\,\text{\textmu{}s}$, the drop is still approaching the surface. The
pressure builds up close to $x=0$. Since the liquid near $x=0$ is decelerated
faster, this creates a region of negative vorticity over $0 \lessapprox x
\lessapprox 250\,\text{\textmu{}m}$. By $t=72.05\,\text{\textmu{}s}$, the thin
layer of gas has formed, and the position of the front from
\autoref{fig:drop-profile}(a) is marked with a small triangle. There is a region of
large positive pressure behind the front, a small region of negative pressure
ahead of it. The contour of zero vorticity follows the front as it moves
outward to $t=96.07\,\text{\textmu{}s}$.

\autoref{fig:drop-profile}(d--f) shows zoomed-in plots of the height profiles in
the thin layer for the three simulations. Behind the front, the height profiles
align on top of each other, and trace out relatively stable envelopes,
appearing as thick blue lines in \autoref{fig:drop-profile}(d--f). In all three
cases the envelopes initially slope downward before curving upward, indicating
the deviation of the liquid--gas interface away from the solid surface, prior to contact between the solid and the liquid. This is the lift-off phenomenon \citep{kolinski2014lift}. Lift-off occurs more quickly for lower viscosities,
and the envelope slopes upward faster. These results are in close agreement
with the experimental results of \citet{kolinski2014lift}.

We now quantify the lift-off time and position. While the envelopes
in \autoref{fig:drop-profile}(d--f) are relatively well defined, the height
profiles shift slightly in over time, make it difficult to precisely define
the moment of lift-off. However, close inspection of
\autoref{fig:drop-profile}(d--f) shows that in all cases, the height profiles have
a leading tip that dips slightly below the envelope that forms. We found this
to be a well-defined feature that occurs universally across all of our
simulations. The tip position can be determined as the global minimum of the
height profile at each time, and provides a way to precisely define
when lift-off occurs, at the moment when the leading tip starts to rise.

\begin{figure}
  \centering
  \includegraphics[width=\linewidth]{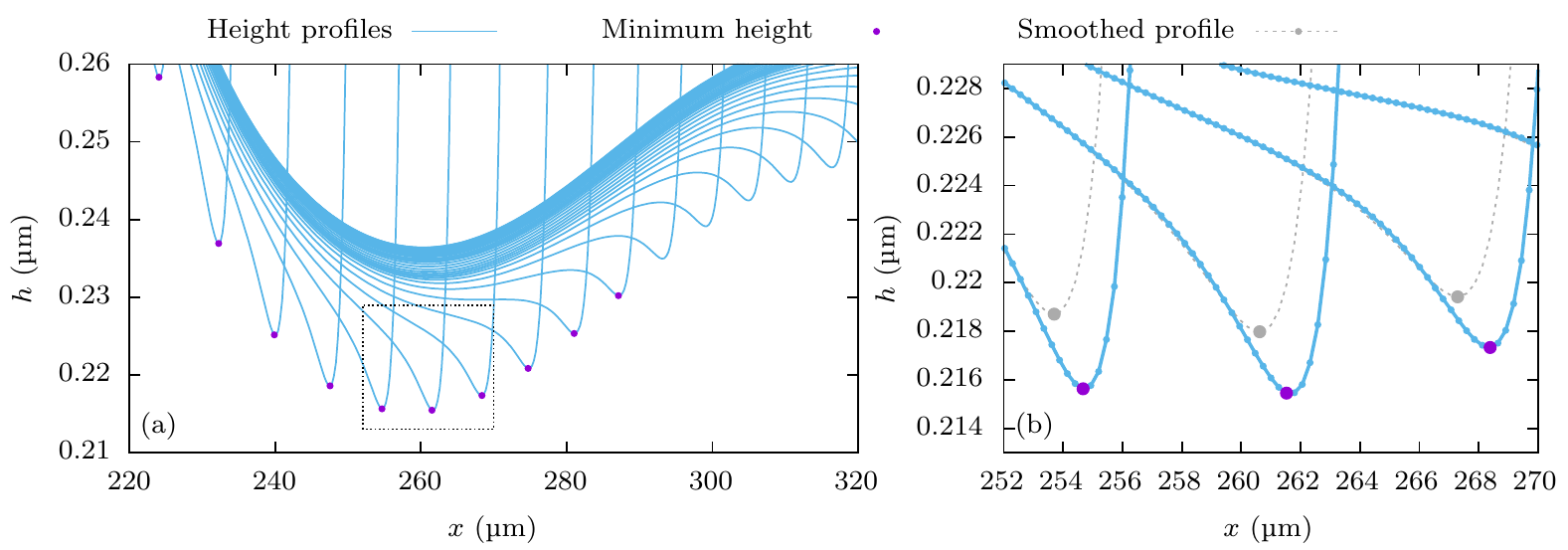}
  \caption{Zoomed-in plots of the height of the gas layer at intervals spaced $1.2009\,\text{\textmu{}s}$ apart for liquid viscosity $\nu_l = \vten$. Panel (a) shows the region marked by the dashed box in \autoref{fig:drop-profile}(d). For each profile, the global minimum, which follows the leading tip, is also plotted on the curves; once the global minimum is no longer at the leading tip, it is no longer plotted. Panel (b) shows the region marked by the dashed box in panel (a). In panel (b), the small blue circles indicate the computational grid. The gray dashed lines show the profiles with Gaussian smoothing applied, and the gray circles show the global minima of the smoothed lines.\label{fig:drop-profiles_zoom}}
\end{figure}

\autoref{fig:drop-profiles_zoom}(a) shows a
further zoomed-in plot of the height profiles during lift-off for the case of
$\nu_l=\vten$. Once the leading tip has risen sufficiently, then it
is no longer the global minimum. However, since lift-off has always occurred by
the time the leading tip is rising, this does not cause any difficulty in
identifying the lift-off time. It is natural to consider whether the leading
tip is a real feature or a numerical artifact that emerges from discretization
error and limited resolution. To address this, \autoref{fig:drop-profiles_zoom}(b)
shows further zoomed-in region, depicting the leading tip in detail. Here, the
blue circles show individual simulation grid points. The grid spacing is
substantially smaller than the width of the leading tip, indicating that it is
a real feature. Further numerical tests of accuracy are provided in Appendix
\ref{app:drop-accur}.

The width of the tip is governed by surface tension. While many of our
simulations use the baseline value of $\sigma = 0.072 \, \text{N}/\text{m}$
from \autoref{tbl:drop-xu-exp-pot}, we also consider reduced surface tension where
the tip becomes sharper. In this case, there can be numerical difficulties with
in identifying the minimum due to per-gridpoint variations the height profile.
To create a scheme for identifying the minimum across all simulations, we
therefore smooth the height profile using a Gaussian kernel with length scale
$1.2\,\text{\textmu{}m}$. This results in a minimal alteration in the leading
tip position (\autoref{fig:drop-profiles_zoom}(b)), but improves robustness when
analyzing the simulations. Numerically, the global minimum is found by
identifying the local minima of the cubic interpolant of the smoothed profile,
and selecting the smallest one.

\begin{figure}
  \centering
  \includegraphics[width=\textwidth]{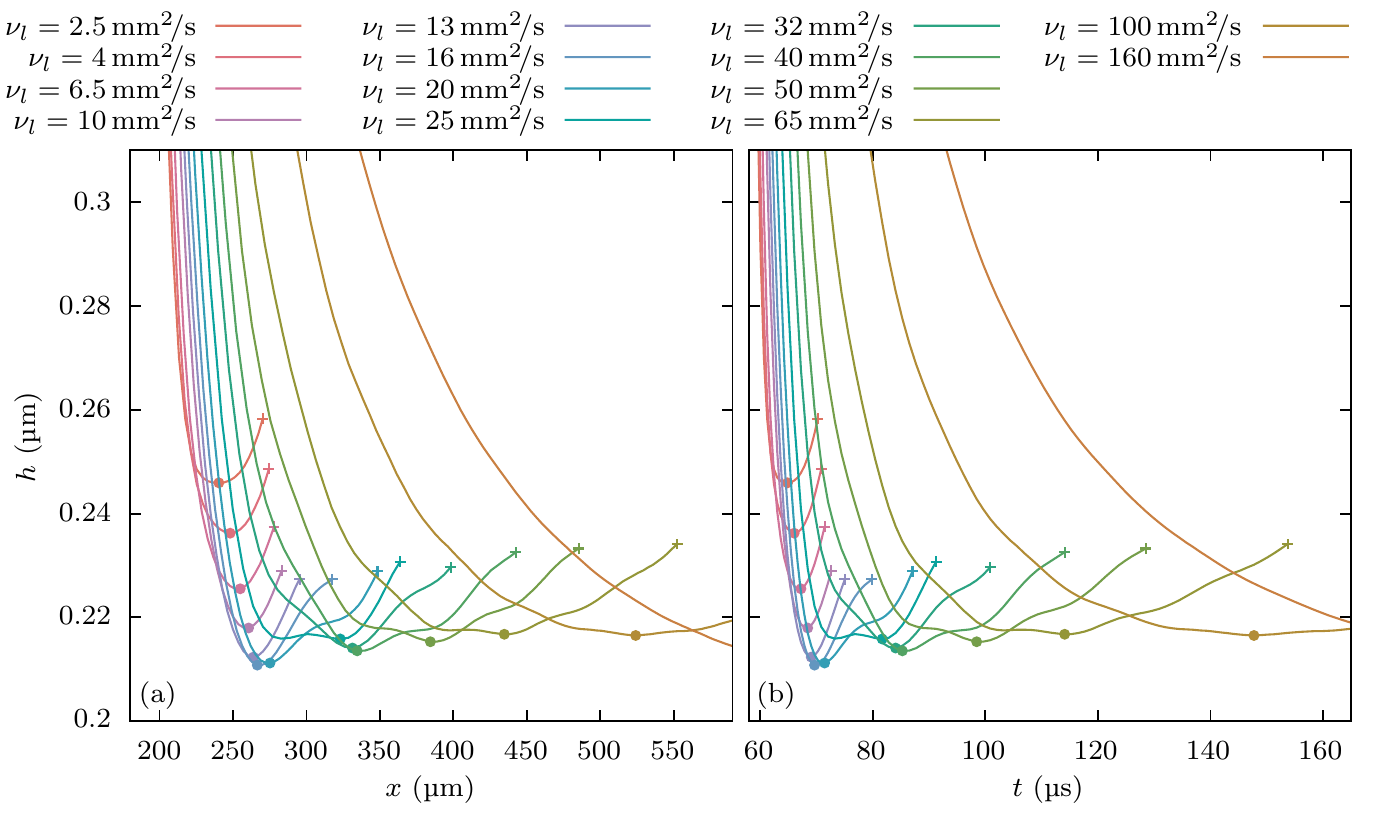}
  \caption{Trajectories of the global minimum of the drop profile in (a) the $(x,h)$ plane, and (b) the $(t,h)$ plane for the baseline parameters with initial drop velocity $V=0.45\,\text{m}/\text{s}$, for a range of different liquid viscosities. For each trajectory, the filled circle indicates where the global minimum reaches its lowest point, which we define as when lift-off occurs. Each cross indicates when the global minimum no longer marks the leading tip.\label{fig:drop-min_traj}}
\end{figure}

\begin{figure}
  \centering
  \includegraphics[width=\textwidth]{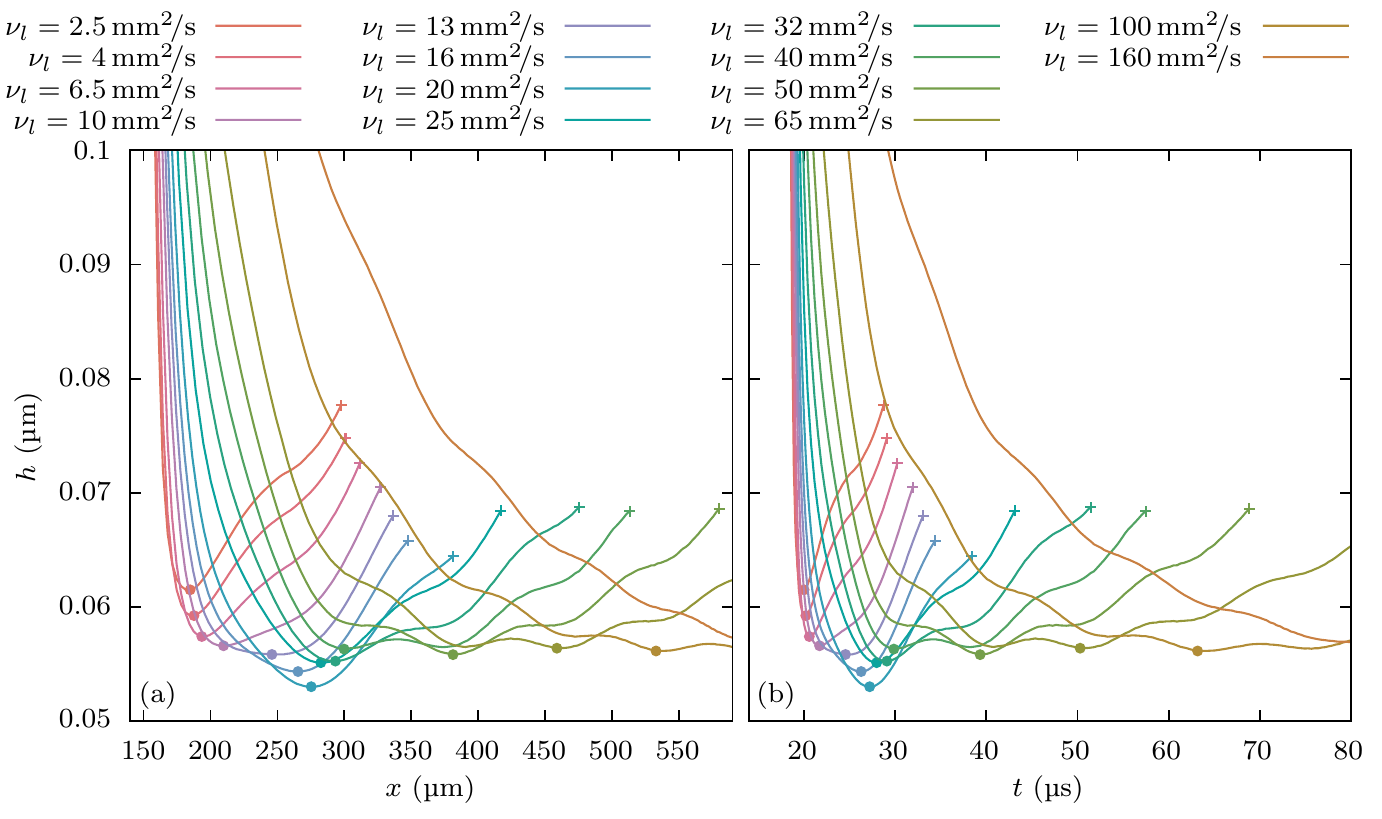}
  \caption{Trajectories of the global minimum of the drop profile in (a) the $(x,h)$ plane, and (b) the $(t,h)$ plane for the baseline parameters with initial drop velocity $V=0.9\,\text{m}/\text{s}$, for a range of different liquid viscosities. For each trajectory, the filled circle indicates where the global minimum reaches its lowest point, which we define as when lift-off occurs. Each cross indicates when the global minimum no longer marks the leading tip.\label{fig:drop-min_traj_v09}}
\end{figure}

\autoref{fig:drop-min_traj} shows the trajectories of the leading tip for a range
of liquid viscosities from $\nu_l=\vtwoptfive$ to $\nu_l=\vhunsixty$.
The moment of lift-off is indicated on each trajectory, by identifying the
global minimum of each curve. Two regimes are visible. For $\nu_l \lessapprox
\vtwenty$, increasing viscosity results in the trajectory reaching a
lower value of $h$, and the lift-off position moving slightly outward. For
$\nu_l \gtrapprox \vtwenty$, increasing viscosity results in the
trajectory reaching a higher value of $h$, and the lift-off position moves
outward substantially. Small undulations in the trajectories are visible for
$\nu_l \gtrapprox \vtwenty$, which arise due to capillary waves in the
height profile. This can have a substantial effect on the measured lift-off
time, depending on which undulation achieves the global minimum. For example,
there is a large difference between $\nu_l = \vtwenty$ and $\nu_l = \vtwentyfive$.
In \autoref{fig:drop-min_traj_v09} we show trajectories for the case when
the initial drop velocity is doubled to $V=0.9\,\text{m}/\text{s}$. The scale
of $h$ is smaller in the plot, but the relative amplitude of the capillary
waves is increased. As a result they have a noticeable effect of lift-off times
at lower viscosities, with a large difference in lift-off time between $\nu_l
= \vten$ and $\nu_l =\vthirteen$.

We now compare to the experimental finding of \citet{kolinski2014lift} that the lift-off time is proportional to
\smash{$\nu_l^{1/2}$}. The lift-off time $\tau$ in this previous work is defined
relative to a time origin of when the drop would reach $h=0$ in the absence of
the surface. Here, since the initial height $h_0$ and velocity $V$ are known,
we compute the time origin as $t_0=h_0/V$. By contrast, \citet{kolinski2014lift} were not able to directly view $h_0$, since the
drop can only be observed once it enters an evanescent field of height
$h_\text{ev}=1\,\text{\textmu{}m}$. Instead, they measured the first position
and time $(h',r',t')$ when the drop enters the evanescent field, and assume
that the drop is undeformed at this position, so that $h' = h_0 - Vt' +
r'^2/(2R^2)$. Hence $h_0$ can be calculated, and therefore the time origin is
$t^\text{alt}_0=h_0/V - t' + r'^2/(2R^2V)$.

\begin{figure}
  \centering
  \includegraphics[width=\textwidth]{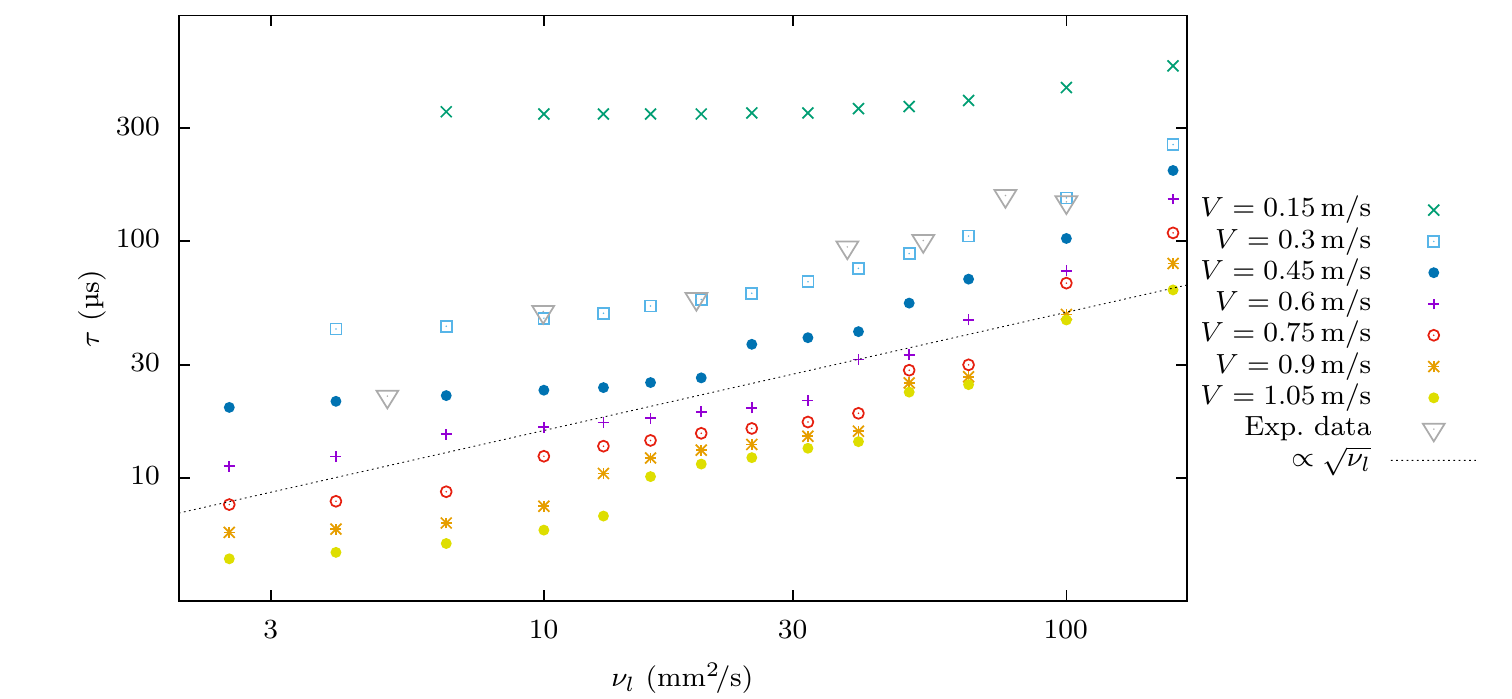}
  \caption{Liftoff time $\tau$ as a function of liquid viscosity $\nu_l$, for a
  range of initial drop velocities $V$. The experimental data is taken
  from Figure 4(a) of the paper by \citet{kolinski2014lift} (which
  used $V=0.45\text{~m}/\text{s}$). Three data points are omitted from the plot
  for low $V$ and low $\nu_l$ due to physical difficulties with running the
  simulation---see \autoref{sec:drop-surf_ten}.\label{fig:drop-lo_vel}}
\end{figure}

\autoref{fig:drop-lo_vel} shows the lift-off times $\tau$ as a function of viscosity
for seven different values of initial velocity $V$. The data points from
\citet{kolinski2014lift} are also plotted. Even though the
experiments of \citet{kolinski2014lift} are performed in a three-dimensional
axisymmetric configuration, there is good quantitative agreement with the
two-dimensional simulation results.

Overall, the results are consistent with the \smash{$\nu_l^{1/2}$} scaling
result, but the additional precision provided by the simulations reveals a more
complicated relationship between $\nu_l$ and $\tau$. For each value of $V$, the
data points exhibit some fluctuations, which are due to the undulations visible
in \autoref{fig:drop-min_traj} where the global minimum defining the lift-off
time may abruptly change. While a \smash{$\nu_l^{1/2}$} scaling appears
consistent with the high impact velocities where $V\gtrapprox
0.75\,\text{m}/\text{s}$, the data for low impact velocities looks better fit
to $\nu_l^\eta$, where there are two different values of $\eta$ with a change
at $\nu_l \approx \vtwenty$. This is also consistent with the two different
types of behavior observed in \autoref{fig:drop-min_traj}.

The slope of data points for small values of $\nu_l$ in \autoref{fig:drop-lo_vel}
is strongly affected by the choice of time origin, and may affect the
conclusions about the relevant exponents. To investigate this further, we
replotted the data using $t_0^\text{alt}$ as the time origin; this resulted in
small shift upwards of the data (since $t_0^\text{alt}<t_0$ in all cases), but
did not affect the overall patterns. As a third approach, we defined a time
origin $t_0^\text{dat}$ directly from the data, by finding the best fit to the
model $\tau_\text{dat} = t-t_0^\text{dat} = \gamma
(\nu_l/\nu_\text{sc})^\alpha$ for the three free parameters $(t_0^\text{dat},
\gamma, \alpha)$. Here $\nu_\text{sc}=\vtwenty$ is chosen as an
arbitrary viscosity scale. Specifically, for the data points $(\nu_{l,k},t_k)$,
we minimized the residual
\begin{equation}
  \label{eqn:drop-resid}
  S(t_0^\text{dat},\gamma,\alpha) =\frac{1}{2} \sum_k \left( \log \gamma + \alpha \log \frac{\nu_{l,k}}{\nu_\text{sc}} - \log(t_k-t_0^\text{dat})\right)^2.
\end{equation}
We used the L-BFGS-B algorithm for bound-constrained nonlinear
optimization~\citep{byrd95}, and enforced $t_0^\text{dat}<0.999 t_\text{min}$
and $\alpha>0$, where $t_\text{min} = \min_k \{t_k\}$. We started the
minimization using multiple initial guesses with $t_0^\text{dat} \in
[0.7t_\text{min},0.9t_\text{min}]$ and $\alpha \in [0.4,1.2]$. For each pair
$(t_0^\text{dat},\alpha)$ the value of $\gamma$ is set so that the bracketed
expression in equation \eqref{eqn:drop-resid} is zero for the $\nu_l=\vtwenty$
data point, so that the initial guess should be close to the minimum of $S$.

\begin{figure}
  \centering
  \includegraphics[width=\textwidth]{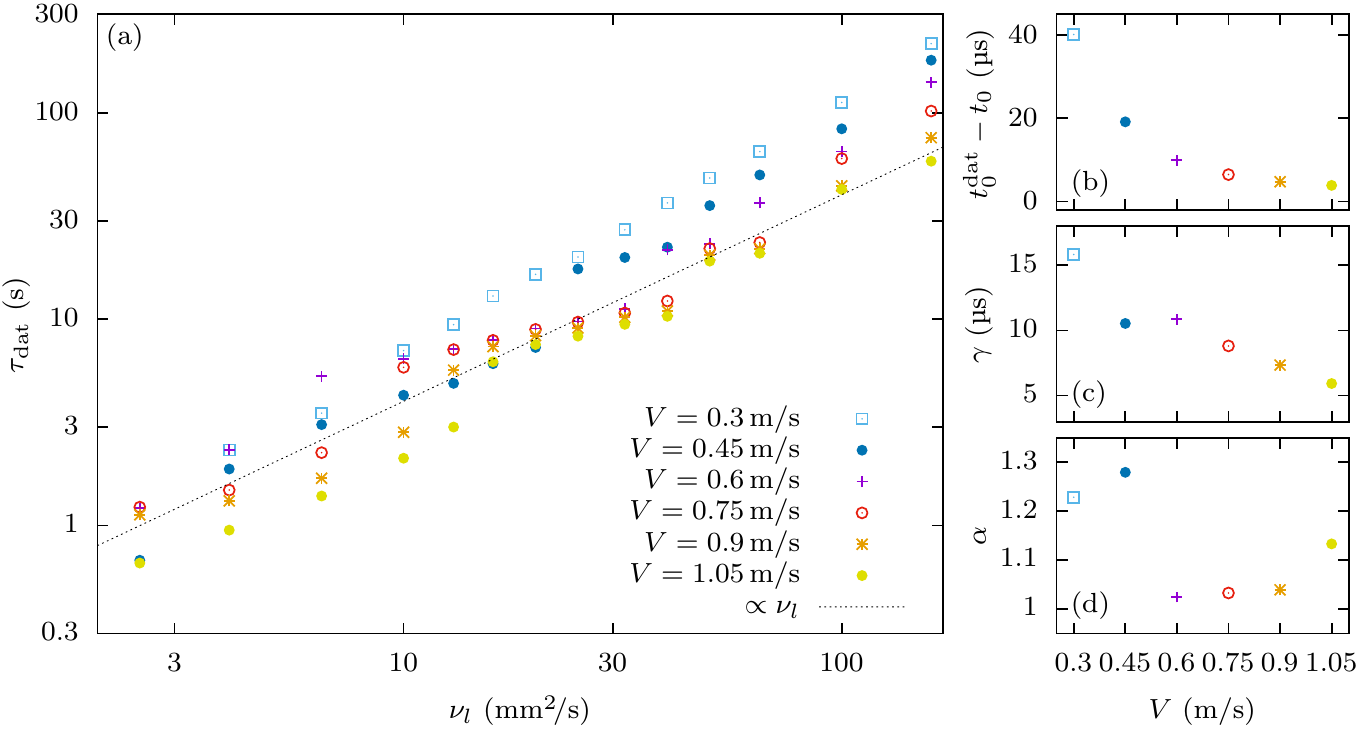}
  \caption{(a) Liftoff time $\tau_\text{dat}$ as a function of liquid viscosity
  $\nu_l$, for a range of initial drop velocities $V$, using the
  alternative time origin definition based on minimizing the three
  parameter residual function in equation \eqref{eqn:drop-resid}. (b--d) best fits
  of the parameters $t_0^\text{dat}- t_0$, $\gamma$, and $\alpha$ for different
  initial drop velocities.\label{fig:drop-lo_vel_dat}}
\end{figure}

\autoref{fig:drop-lo_vel_dat} shows a replotting of the data relative to this
time origin. The data for $V=0.15\,\text{m}/\text{s}$ is omitted from this plot
since the lift-off times are non-monotonic for the smallest values of $\nu_l$.
For each other value of $V$, the minimization procedure converges to a single
unique solution for all initial guesses. With this definition of time origin,
all of the data is more consistent with a linear scaling relationship as
opposed to the square root relationship in \autoref{fig:drop-lo_vel}. Panels (b),
(c), and (d) show the values of the fitted parameters, and demonstrate that
$\alpha \in [1.02,1.28]$ in all cases. The average residual over the six
different values of $V$ is $\bar{S}=0.2536$. If the exponent is constrained to
\smash{$\alpha=\tfrac{1}{2}$} to match the exponent of
\citet{kolinski2014lift} then the average residual increases to
$\bar{S}=0.3855$ and the data points exhibit an upward curve a log--log plot
that systematically deviates from a power law. Constraining the exponent to
$\alpha=1$ results in $\bar{S}=0.2805$ and a better fit to the model that is
similar to the three-parameter fit shown in \autoref{fig:drop-lo_vel_dat}. See
the \autoref{sec:drop-app-model-fits} for additional discussion and parameter fits.
These results highlight that any theoretical analysis of the relationship
between $\tau$ and $\nu_l$ would need to take into account the sensitivity of
defining the time origin.

\begin{figure}
  \centering
  \includegraphics[width=\textwidth]{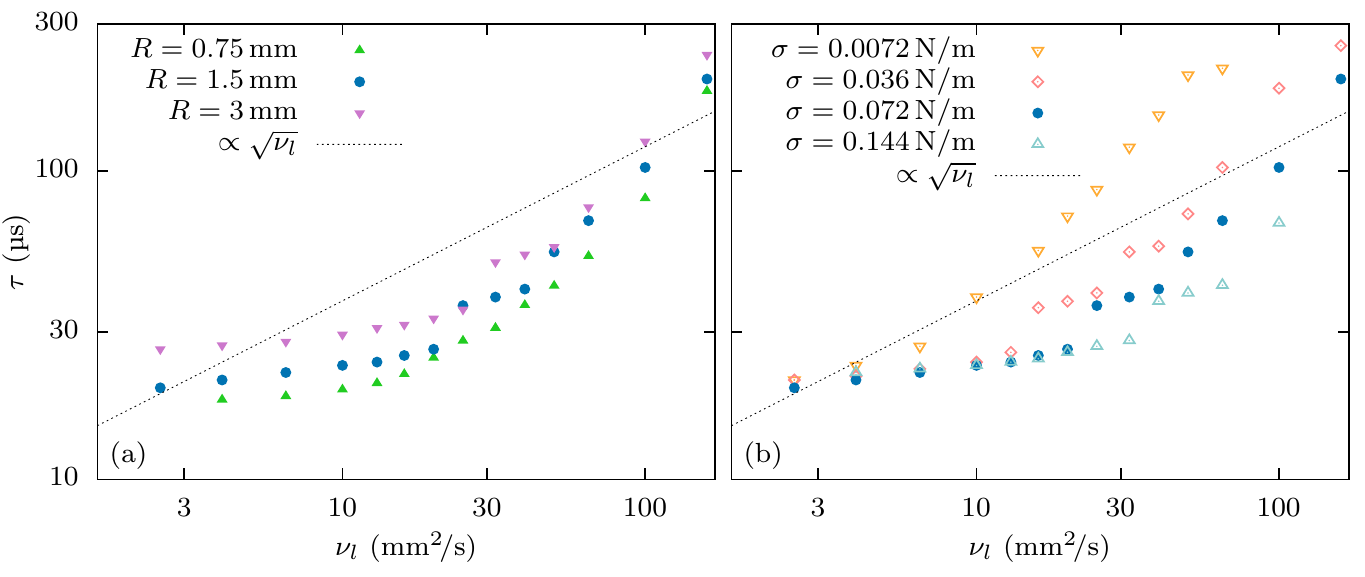}
  \caption{Lift-off time as a function of liquid viscosity, for a range of (a) different drop radii, and (b) different surface tension values. The data point for $(\nu_l,R)=(\vtwoptfive,0.75\,\text{mm})$ is omitted due to physical difficulties with running the simulation (\autoref{sec:drop-surf_ten}). The data points for $\sigma=0.0072\,\text{N}/\text{m}$ and $\nu_l\ge\vhun$ are omitted because lift-off does not occur over the simulation duration.\label{fig:drop-lo_sigma_rad}}
\end{figure}

\autoref{fig:drop-lo_sigma_rad}(a) shows lift-off times (relative to $t_0$) for
for three different drop radii: the original value
of $R=1.5\,\text{mm}$, and as well as half and double this value. Increasing the
radius increases the lift-off time, similar to the effect of lowering velocity
in \autoref{fig:drop-lo_vel}.

\subsection{The role of surface tension}
\label{sec:drop-surf_ten}
The simulations allow us to change the surface tension in ways that would
be difficult to do experimentally, to investigate the importance of this
physical effect. Changing the surface tension allows us to suppress or
accentuate the capillary waves in the liquid--gas interface, as shown in
\autoref{fig:drop-min_traj}, to investigate the effect of surface tension on
the phenomenon of lift-off. Following the same procedure as in
\autoref{sec:drop-visc_liftoff}, we calculated the lift-off times for
$\sigma=0.0072\,\text{N}/\text{m}$, $\sigma = 0.036\,\text{N}/\text{m}$ and the
$\sigma = 0.144\,\text{N}/\text{m}$, corresponding to a tenth, half, and double
the usual surface tension values, respectively.
\autoref{fig:drop-lo_sigma_rad}(b) shows that as surface tension is reduced,
the lift-off times increase markedly. For the case of
$\sigma=0.0072\,\text{N}/\text{m}$ lift-off is eliminated for large values of
$\nu_l$, with the leading tip continuing to decrease over the course of the
simulation.

\begin{figure}
  \centering
  \includegraphics[width=\linewidth]{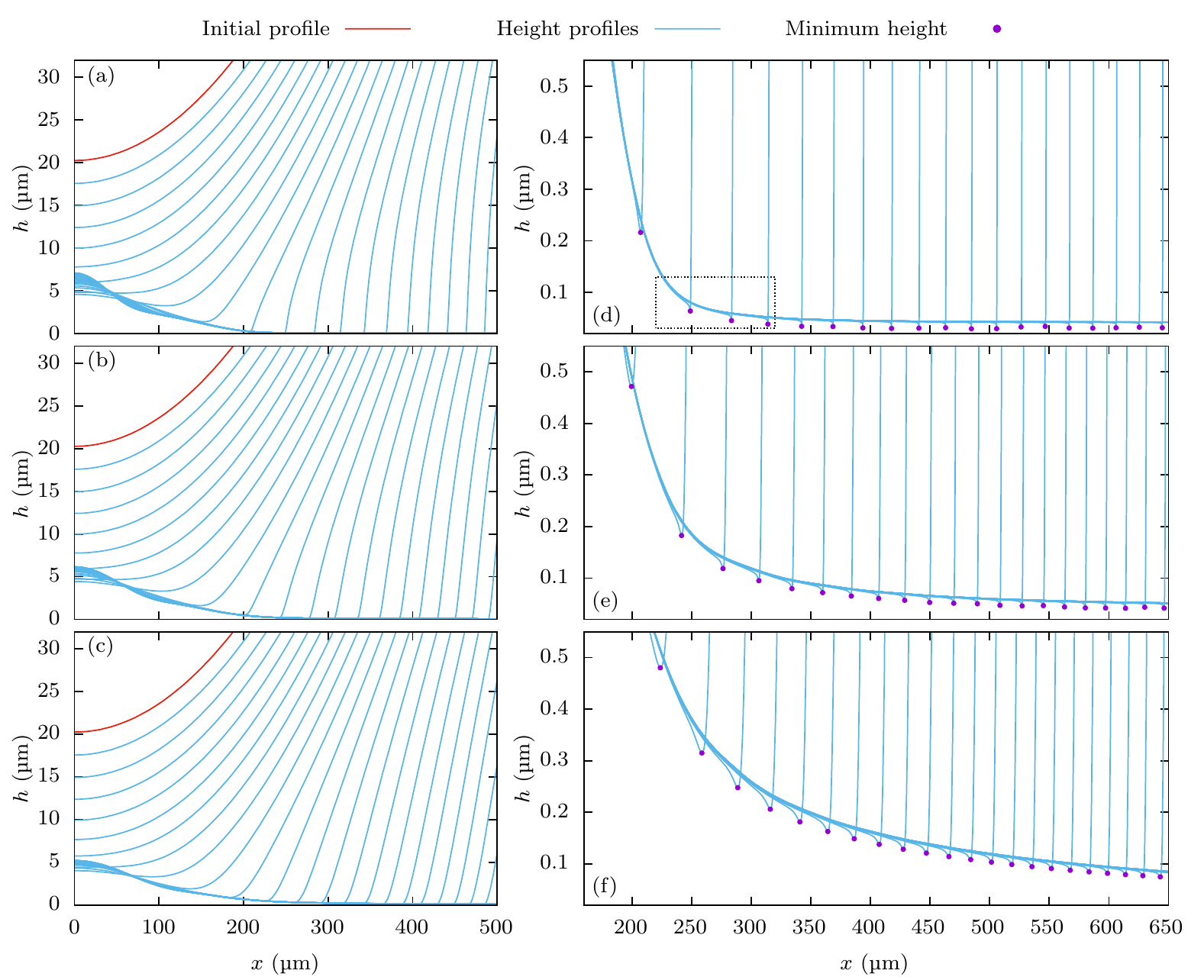}
\caption{Profiles of the height of the gas layer at intervals spaced $6.004\,\text{\textmu{}s}$ apart for liquid viscosities of (a) $\nu_l = \vten$, (b) $\nu_l=\vthirtytwo$, and (c) $\nu_l=\vhun$ using zero surface tension. Panels (d), (e), and (f) show the same data as (a), (b), and (c), respectively, but with a smaller range of $h$ to highlight that no lift-off occurs in this case. For each profile, the global minimum, which follows the leading tip, is also plotted on the curves. The dashed box in panel (d) marks a further zoomed-in region shown in \autoref{fig:drop-profiles_st_zoom}.  \label{fig:drop-profiles_st}}
\end{figure}

\begin{figure}
  \centering
  \includegraphics[width=\linewidth]{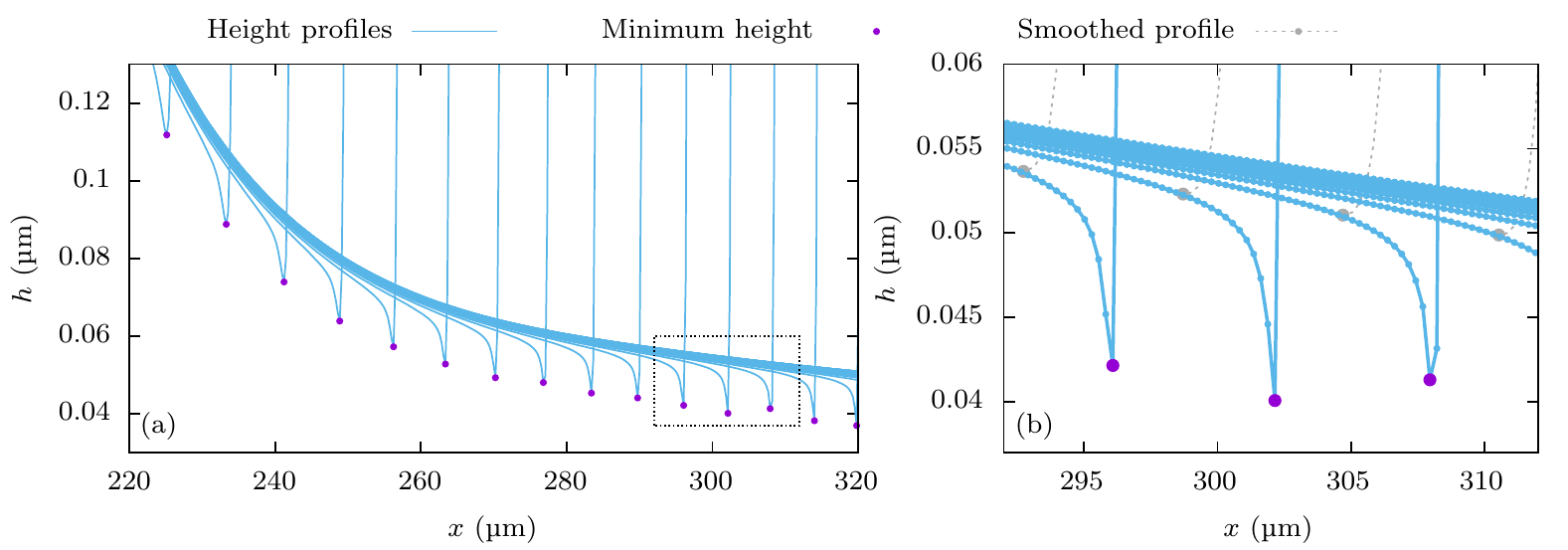}
  \caption{Zoomed-in plots of the height of the gas layer at intervals spaced $1.2009\,\text{\textmu{}s}$ apart for liquid viscosities of $\nu_l = \vten$ using zero surface tension. Panel (a) shows the region marked by the dashed box in \autoref{fig:drop-profiles_st}(d). For each profile, the global minimum, which follows the leading tip, is also plotted on the curves. Panel (b) shows the region marked by the dashed box in panel (a). In panel (b), the small blue circles indicate the computational grid. The gray dashed lines show the profiles with Gaussian smoothing applied, and the gray circles show the global minima of the smoothed lines.\label{fig:drop-profiles_st_zoom}}
\end{figure}

\autoref{fig:drop-lo_sigma_rad}(b) strongly suggests that surface tension is
important in creating lift-off. Reducing from $\sigma =
0.036\,\text{N}/\text{m}$ to $\sigma=0.0072\,\text{N}/\text{m}$ almost doubles
the lift-off times in most cases, and one can ask whether lift-off will be
completely eliminated in the limit as surface tension vanishes. To examine
this, we ran a sequence of simulations with $\sigma=0$, with several
representative examples for $\nu_l=\vten$, $\nu_l = \vthirtytwo$,
and $\nu_l=\vhun$ shown in \autoref{fig:drop-profiles_st}. This is a
difficult limit to probe in our simulations, since as discussed for
\autoref{fig:drop-profiles_zoom}, surface tension regularizes the leading tip.
\autoref{fig:drop-profiles_st_zoom} shows close-ups of the profiles for
$\nu_l=\vten$, indicating that leading tip becomes as sharp as a
single grid point, and the minima of the profiles can fluctuate
non-monotonically depending on exactly how the tip aligns with the
computational grid. However, in this case the profile smoothing procedure
introduced in \autoref{sec:drop-visc_liftoff} is sufficient to extract smooth
leading tip trajectories.

The very sharp leading tip causes another difficulty in the simulations: as the
tip is advected across the computational grid, there will be an effective
numerical diffusion, which will act as though a small surface tension has been
imposed. This is an important issue since \autoref{fig:drop-lo_sigma_rad}(b) already
confirms that small surface tensions can considerably alter the behavior. To
test this, we compared simulations with the baseline parameters to those on a
finer grids (Appendix \ref{app:drop-accur}). Unlike the case for finite surface
tension, the simulations on a finer grids are noticeably different, with the
profiles reaching lower heights and the leading tip not curving up as rapidly.
Since liquid viscosity also regularizes the evolution of the height
profile, these discrepancies are more significant when $\nu_l$ is small.

\begin{figure}
  \centering
  \includegraphics[width=0.95\textwidth]{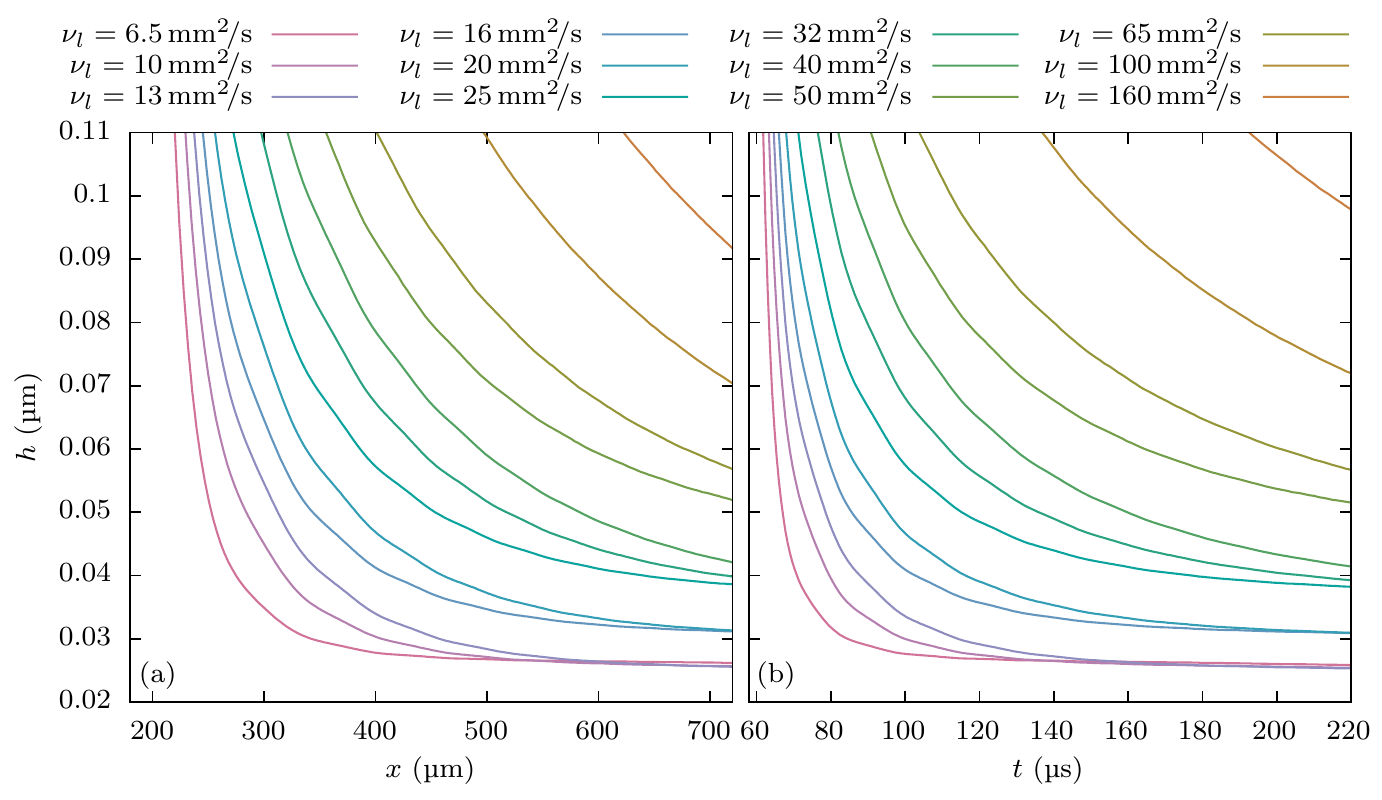}
  \caption{Trajectories of the global minimum of the drop profile in (a) the $(x,h)$ plane, and (b) the $(t,h)$ plane for a range of different liquid viscosities, using the baseline parameters and zero surface tension. Simulations with $\vthirteen <\nu_l \le \vforty$ use a grid of size $8192\times 1536$, and simulations with $\nu_l \le \vthirteen$ use a grid of size $12288 \times 2304$.\label{fig:drop-min_traj_st}}
\end{figure}

We calculated the trajectories of the leading tip for a range of different
liquid viscosities $\nu_l$. For $\vthirteen <\nu_l \le \vforty$ we switched to a
larger computational grid of size $8192\times 1536$, and for $\nu_l \le \vthirteen$
we switched to a very large grid of size $12288 \times 2304$. For $\nu_l \le 4
\vfour$ we were not able to simulate on a large enough grid to adequately
resolve the numerical diffusion, and hence results are omitted.
\autoref{fig:drop-min_traj_st} shows the trajectories in both the $(x,h)$ and
$(t,h)$ planes. In all cases, the leading tips continue to decrease. While our
results only examine one particular set of physical parameters (from
\autoref{tbl:drop-xu-exp-pot}) our results strongly suggest that surface
tension is required for lift-off to occur.

\begin{figure}
  \centering
  \includegraphics[width=\linewidth]{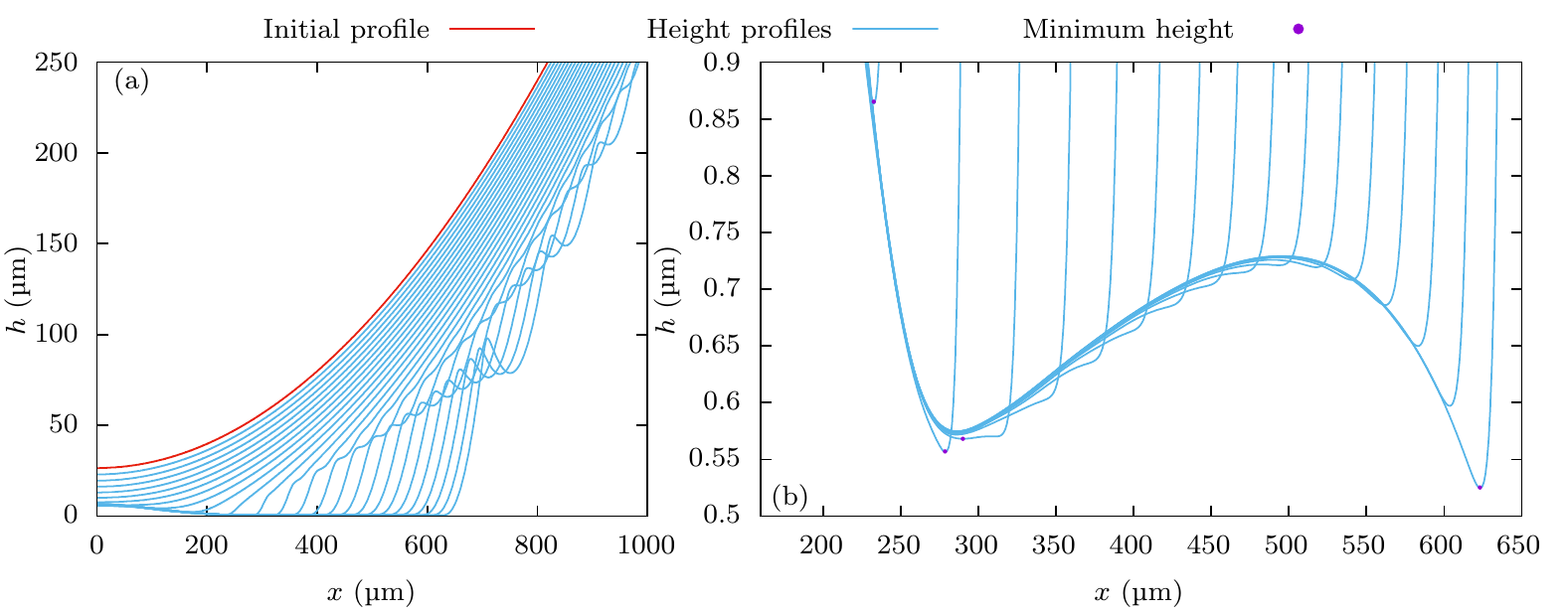}
  \caption{(a) Profiles of the height of the gas layer at intervals spaced $11.80\,\text{\textmu{}s}$ apart for liquid viscosity of $\nu_l=\vsixptfive$ and initial drop velocity of $V=0.3\,\text{m}/\text{s}$, showing the development of capillary waves. (b) Zoomed-in plot of the same data. The global minima of the curves are plotted when they indicate the leading tip.\label{fig:drop-profiles_cw}}
\end{figure}

We also found a regime where the effect of surface tension can qualitatively
affect the lift-off behavior. \autoref{fig:drop-profiles_cw}(a) shows the height
profiles for a simulation with the baseline value of surface tension of
$\sigma=0.072\,\text{N}/\text{m}$ in the regime of low initial velocity,
$V=0.3\,\text{m}/\text{s}$ and low viscosity, $\nu_l = \vsixptfive$. In
this regime, prominent capillary waves are generated outside the thin gas
layer, which grow larger as time progresses. \autoref{fig:drop-profiles_cw}(b)
shows a zoomed-in plot of the profiles in the thin gas layer. Lift-off appears
to occur at $x \approx 280 \, \text{\textmu{}m}$ and the leading tip starts to
rise. However the influence of the capillary wave causes the tip to move
downward again, ultimately dipping below the previous minimum height. It is
possible that the tip may move upward again at a later point, but we were
unable to track the behavior further. The sharp features visible in
\autoref{fig:drop-profiles_cw}(a) (e.g.~at
$(x,h)\approx(700\,\text{\textmu{}m},100\,\text{\textmu{}m})$) cause the
simulation to terminate early, since the Newton--Raphson iterations can no
longer be solved to an acceptable level of accuracy.

\section{Conclusion}
In this paper we investigated the viscous effects in the early stages of drop impact on a surface. We coupled a Navier--Stokes solver to model flow in the interior of the drop with a partial differential equation to model the pressure and height in the thin gas layer between the drop and the surface. We demonstrated that our simulations are consistent with previous work using potential flow theory where flow in the liquid is assumed incompressible~\citep{mandre2009precursors, mani2010events, mandre2012mechanism}. However, our simulations allow us to go beyond this previous work and investigate viscous effects. We showed that at low initial drop velocities, viscosity plays a weak role in the deceleration of the drop, and the height $H^*$ at which it reaches a stagnation point.

Using our simulations, we are able to recreate the lift-off phenomenon that was experimentally reported by \citet{kolinski2014lift}. We have therefore demonstrated that with the reduced model described in \autoref{sec:drop-model}, our simulations are able to capture lift-off. Our numerical results for the lift-off time $\tau$ are consistent with the \smash{$\nu_l^{1/2}$} scaling relationship found by \citet{kolinski2014lift}. However, the additional precision afforded by the simulations indicates that the precise relationship between $\tau$ and $\nu_l$ is more complex, due to the effects of capillary waves and the sensitivity of the results to the definition of the time origin. The simulation allows us to probe conditions that would be difficult to observe experimentally. Using this capability, our results provide strong evidence that surface tension is necessary for lift-off to occur.

There are a number of possible next steps. The simulations provide a detailed view of the lift-off phenomenon, and all aspects (e.g.~stress, velocity, vorticity) can be calculated. The results may therefore guide theoretical analyses of lift-off, and may allow scaling relationships similar to those presented by \citet{mandre2009precursors,mandre2012mechanism} and \citet{mani2010events} to be derived. As part of our study, we have released a complete high-performance open source code that can examine lift-off across a wide range of configurations.

We opted to use a fixed-grid simulation for simplicity in coupling the liquid domain and gas layer together, and as described in Appendices~\ref{app:drop-perf} and \ref{app:drop-accur} we are able to obtain accurate simulation results in a reasonable timeframe. However, it is likely that the behavior at the bottom of the liquid domain, close to the liquid--gas interface and near the leading tip, dominates. Because of this, the simulation is a good candidate for use with mesh refinement~\citep{berger84,guittet15a}, where the liquid--gas interface and the region around the leading tip could use a finer mesh. This could result in substantial computational savings, although the liquid--gas coupling would become more complicated and numerical errors would be harder to quantify.

The simulation could also be generalized to use cylindrical axisymmetric coordinates. The overall numerical approach would stay the same, but additional radial factors would have to be incorporated throughout the simulation. The Navier--Stokes solver that we employ has already been demonstrated to work in axisymmetric coordinates~\citep{yu2003coupled,yu07}, but the routines that involve the gas layer would require modifications. For example, the kernel in equation \eqref{eqn:drop-p-int} would need to be modified. While the current simulation is already in good agreement with the experimental results of \citet{kolinski2014lift}, an axisymmetric solver would allow for a near-perfect comparison. This would, for example, help elucidate further the precise relationship between lift-off time $\tau$ and liquid viscosity $\nu_l$.

Experiments by Thoroddsen and coworkers have used interferometry to examine the evolution of the gas layer across a full two-dimensional surface~\citep{langley17}. Their results show a number of interesting effects beyond the scope of our current model, such as a breakage of rotational symmetry and the formation of ruptures in the air film~\citep{langley17,li17}. They have also examined the case of nano-rough surfaces~\citep{langley18}. To connect with this work, our model could be generalized to a full three-dimensional simulation of the liquid and two-dimensional simulation of the gas-layer. This would be substantially more computationally challenging, and would be a good candidate for using parallel computing and adaptive mesh refinement~\citep{zhang19}.
% END BODY

\backsection[Funding]{C.~H.~R.~was partially supported by the National Science Foundation under Grant No.~DMS-1753203. C.~H.~R.~was partially supported by the Applied Mathematics Program of the U.S.~DOE Office of Science Advanced Scientific Computing Research under Contract No.~DE-AC02-05CH11231.}

\backsection[Declaration of interests]{The authors report no conflict of interest.}

\backsection[Data availability statement]{The C++ code to generate all of these results is provided in the GitHub repository at \url{https://github.com/chr1shr/vdropimpact}. The code makes use of the TGMG library, which is provided in a separate GitHub repository at \url{https://github.com/chr1shr/tgmg}.}

\backsection[Author ORCID]{C.~H.~Rycroft,  https://orcid.org/0000-0003-4677-6990; S.~Mishra,  https://orcid.org/0000-0002-9487-5531; S.~M.~Rubinstein, https://orcid.org/0000-0002-2897-2579}

\appendix

\section{Calculations for the governing equation in the gas layer}
\subsection{Derivation of the gas layer pressure update equation}
\label{app:drop-gas_eq_deriv}
Substituting for $\rho_g$ from the equation of state, equation \eqref{eqn:drop-adia}, into the lubrication equation for the pressure in the gas layer, equation \eqref{eqn:drop-lubr}, we get
\begin{equation}
12 \mu_g \left(\left(\frac{\rho_0}{p_0^{1/\gamma}} p_g^{1/\gamma} \right)h\right)_t = \left(\left(\frac{\rho_0}{p_0^{1/\gamma}} p_g^{1/\gamma} \right) h^3 p_{g,x}\right)_x.
\end{equation}
Dividing both sides by ${\rho_0}/{p_0^{1/\gamma}}$,
\begin{equation}
12 \mu_g \left( p_g^{1/\gamma} h\right)_t = \left(p_g^{1/\gamma} h^3 p_{g,x}\right)_x.
\end{equation}
Using the product rule to expand the brackets,
\begin{equation}
12 \mu_g \left( \frac{1}{\gamma}p_g^{1/\gamma-1} p_{g,t}h
+ p_g^{1/\gamma} h_t \right)
= \frac{1}{\gamma}p_g^{1/\gamma-1}p_{g,x} h^3 p_{g,x}
+p_g^{1/\gamma} 3 h^2 h_x p_{g,x}
+p_g^{1/\gamma} h^3 p_{g,xx}.
\end{equation}
Dividing both sides by $p_g^{1/\gamma-1}$,
\begin{equation}
12 \mu \left( \frac{1}{\gamma} p_{g,t}h
+ p_g h_t \right)
= \frac{1}{\gamma} p_{g,x} h^3 p_{g,x}
+p_g 3 h^2 h_x p_{g,x}
+p_g h^3 p_{g,xx}.
\end{equation}
Dropping the subscript $g$ for gas yields
\begin{equation}
12 \mu \left( \frac{1}{\gamma} p_{t}h
+ p h_t \right)
= \frac{1}{\gamma} p_{x} h^3 p_{x}
+p 3 h^2 h_x p_{x}
+p h^3 p_{xx},
\end{equation}
which can be rearranged to give equation \eqref{eqn:drop-gas-rewrite}.

\subsection{Numerical solution of the gas layer pressure}
\label{app:drop-gas_eq_numerics}
We now describe how to solve equation \eqref{eqn:drop-gas-fd} in the main text for
updating the gas layer pressure using the Newton--Raphson method. Let $P^k$ be
a vector containing the pressure estimates at the $k$th Newton--Raphson
iteration. As an initial estimate, we set $P^0$ to be the pressure values from
the $n$th timestep. We then write \eqref{eqn:drop-gas-fd} as a nonlinear system
$F(P)=0$, where the $i$th component of $F$ is given by the
$(\text{L.H.S.}-\text{R.H.S.})$ of equation \eqref{eqn:drop-gas-fd} evaluated at the
$i$th gridpoint. An improved estimate $P^{k+1}$ for the pressure is given in
terms of $P^k$ by
\begin{equation}
  \label{eqn:drop-newt_raph}
J_F(P^k) \left(P^{k+1} - P^k\right) = -F(P^k),
\end{equation}
where $J_F(P^k)$ is the Jacobian of $F$, which has components
\begin{equation}
J^{k+1}_{F,ij} = \frac{\partial F_i}{\partial P^{k+1}_j}.
\label{eqn:drop-p-jac}
\end{equation}
Since the finite-difference stencils in equation \eqref{eqn:drop-gas-fd} only involve
adjacent grid points, $J_F$ is a tridiagonal system. It can be written as
\begin{equation}
  J_F = \left(
    \begin{array}{ccccc}
      b_0 & c_0 & & & \\
      a_1 & b_1 & c_1 & & \\
          & a_2 & b_2 & c_2 & \\
          &     & \ddots & \ddots & \ddots
    \end{array}
  \right),
\end{equation}
where the terms are given by
\begin{align}
  a_i = \frac{\partial F}{\partial p_{i-1}^{n+1}} & =
  \frac{1}{\gamma} \frac{\bar{h}^3}{\Delta x} \left(\frac{p^{n+1}_{i+1}-p^{n+1}_{i-1}}{2\Delta x}\right)
  - \frac{3 \bar{h}^2 \bar{h}_x }{2 \Delta x} p^{n+1}_i
  + \frac{\bar{h}^3}{\Delta x^2}{p^{n+1}_i}, \label{eqn:drop-newt_a} \\
  b_i = \frac{\partial F}{\partial p_{i}^{n+1}} &=
  -12\frac{\mu}{\gamma}\frac{\bar{h}}{\Delta t} (1) - 12 \mu \bar{h}_t
-\left[
  \frac{3 \bar{h}^2 \bar{h}_x }{2 \Delta x}  \left({p^{n+1}_{i+1}-p^{n+1}_{i-1}}\right)
\right. \nonumber \\
&\phantom{=} \left.
  + \frac{\bar{h}^3}{\Delta x^2} \left(p_{i+1}^{n+1} - 2 p_{i}^{n+1} + p_{i-1}^{n+1}\right)
  + \frac{\bar{h}^3}{\Delta x^2} p_{i}^{n+1}(-2)
\right],    \\
  c_i = \frac{\partial F}{\partial p_{i+1}^{n+1}} &=
  -\frac{1}{\gamma} \frac{\bar{h}^3}{\Delta x} \left(\frac{p^{n+1}_{i+1}-p^{n+1}_{i-1}}{2 \Delta x}\right)
  -  \frac{3 \bar{h}^2 \bar{h}_x }{2 \Delta x} p^{n+1}_i
  - \frac{\bar{h}^3}{\Delta x^2}{p^{n+1}_i}. \label{eqn:drop-newt_c}
\end{align}
Solving equation \eqref{eqn:drop-newt_raph} can be done efficiently using LAPACK's
tridiagonal solver \texttt{dgtsv}~\citep{anderson99}. In most cases, since the
pressure does not change by a large amount per timestep, fewer than five
iterations are required in order to achieve numerical convergence.

\section{Tests of the simulation performance}
\label{app:drop-perf}
\autoref{tbl:drop-perf_stats} contains statistics about the performance of the
code for several simulations that were referenced in the main text. All
tests were run using ten threads on an Ubuntu Linux computer with a ten-core
2.8\,GHz Intel Core i9-10900 CPU, and the code was compiled with GCC version
10.3.

The initial dynamics simulations only need to capture the large-scale
deformation of the drop, and can therefore be run on relatively coarse
computational grids. Consequently, a typical simulation takes approximately
40\,min of wall clock time to run. A large portion of the computation time is
spent on solving the linear systems with the multigrid method. This should be
expected since the multigrid V-cycles involve repeated scans over the entire
grid. Collectively, the four multigrid solves take up 44\% of the total
computation time. The MAC and FEM linear systems take slightly more time and
V-cycles, since apart from where Dirichlet boundary conditions are applied,
these linear systems are only weakly diagonally dominant. By contrast, the
linear systems for the implicit viscous term are strictly diagonally dominant.
We note that the linear systems for the $u$ and $v$ velocity components are
slightly different, since at $x=0$ we apply different boundary conditions,
$(u,v_x)=(0,0)$. Hence it is not possible to vectorize this system, and there
is no substantial computational advantage over solving for the updates to $u$
and $v$ separately.

Calculating the boundary conditions according to \eqref{eqn:drop-p-int} requires
applying Simpson's rule along the $M$ gridpoints on the bottom edge. This must
be done for the top edge of length $M$, and the right edge of length $N$,
requiring $\mathcal{O}(M(M+N))$ work in total. This is sizable amount of work
and takes 3.4\% of the total computation time. Even though the gas layer
involves several Newton steps and tridiagonal matrix solves, it only needs
$\mathcal{O}(M)$ work and therefore takes up a minimal amount of the total
computation time.

\autoref{tbl:drop-perf_stats} also contains performance information for two
lift-off simulations, which are run on larger computational grids to obtain
high accuracy in the height of the gas layer. The wall clock time per timestep
increases from 81~ms to 1.041~s. Based on linear scaling with gridpoints, we
would expect 81~ms to increase to
\smash{$(81\text{~ms})\tfrac{5120\times960}{2048\times256} = 760\text{~ms}$}.
Thus the performance is comparable, but slightly worse than, linear scaling,
which may be due to reduced cache efficiency for a larger grid. Overall,
the percentages spent on the different parts of the simulation are comparable,
although the fraction spent on multigrid V-cycles increases slightly, and the
fraction spent on the gas layer (which scales like $\mathcal{O}(M)$) decreases.
In particular, more V-cycles are required to handle the implicit viscosity
linear system when $\nu_l$ is larger.

\begin{table}
  \centering
  \small
  \begin{tabular}{llll}
    & \shortstack{Initial dynamics, \\ $\nu_l = \vten$}
    & \shortstack{Lift-off, \\ $\nu_l = \vten$}
    & \shortstack{Lift-off, \\ $\nu_l = \vhun$} \\
    \hline
    Grid size & $2048\times 256$ & $5120\times 960$ & $5120\times 960$ \\
    Total WC time & 0.641~h & 20.6~h & 29.8~h\\
    Total timesteps & 28500 & 71250 & 95000 \\
    WC time $/$ timestep & 81~ms & 1041~ms & 1130~ms \\
    \hline
%    WC fraction on output & 0.0011\% & $9.45\times10^{-5}$\% & \\
    WC fraction on gas layer & 0.25\% & 0.047\% & 0.045\% \\
    WC fraction on BCs & 3.39\% & 1.61\% & 1.49\% \\
    WC fraction on MAC solve & 11.21\% & 13.05\% & 12.09\% \\
    WC fraction on FEM solve & 12.77\% & 14.31\% & 14.36\% \\
    WC fraction on viscous solve & 20.11\% & 26.67\% & 31.12\% \\
    \hline
    Mean~MAC V-cycles & 5 & 4.4 & 4.5 \\
    Mean~FEM V-cycles & 5.6 & 4.9 & 5.3 \\
    Mean~viscosity V-cycles & 3.7 & 4.3 & 5.5 \\
  \end{tabular}
  \caption{Performance statistics for several different simulations, compiled
  using GCC 10.3 on an Ubuntu Linux computer with an 2.8\,GHz Intel Core
  i9-10900 CPU. Ten threads were used, and all simulations use the baseline
  parameter choices in \autoref{tbl:drop-xu-exp-pot} \& \ref{tbl:drop-sim_params}
  unless otherwise noted. The wall clock (WC) time is reported for each test,
  and the fraction of time on major components, such as the computation of
  boundary conditions (BCs), solving the marker-and-cell (MAC) linear system,
  and solving finite-element method (FEM) linear system, are
  reported.\label{tbl:drop-perf_stats}}
\end{table}

\begin{figure}
  \centering
  \includegraphics[width=\linewidth]{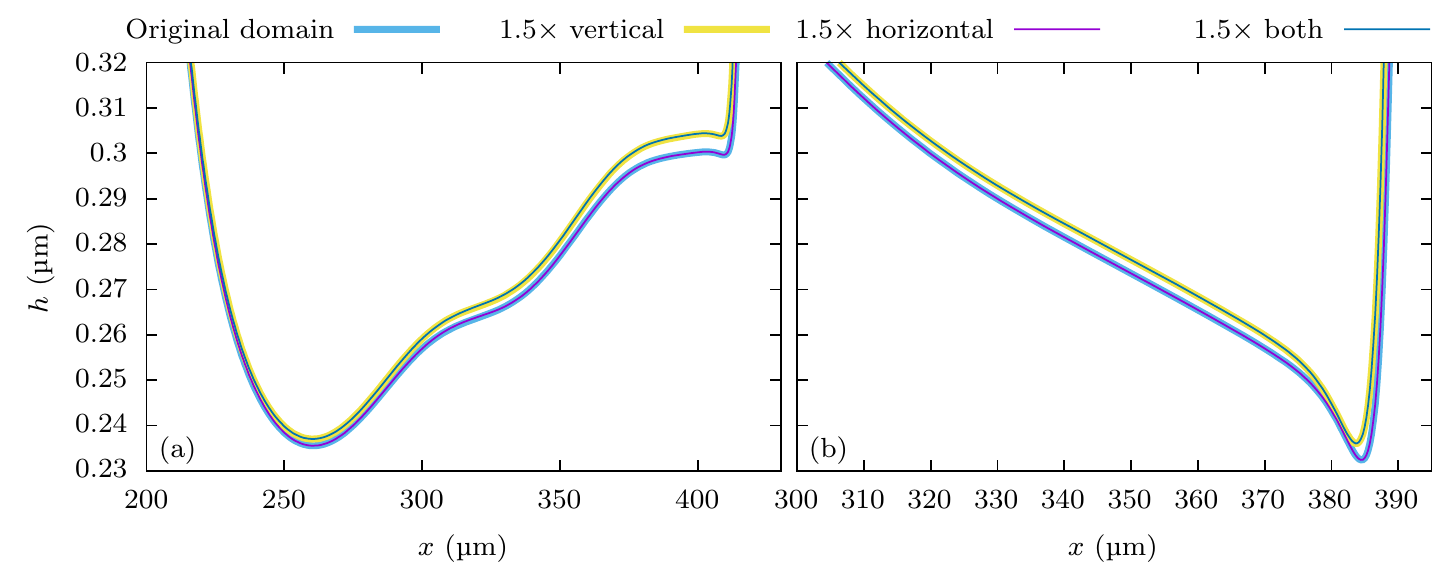}
  \caption{Plots of the height profiles at $t=102.07\,\text{\textmu{}s}$ for
  simulations with (a) liquid viscosity $\nu_l=\vten$ and (b) liquid
  viscosity $\nu_l=\vhun$. Baseline parameters are used although the
  original domain size for both values of $\nu_l$ uses $\tilde{L}=30$ and
  $\beta=\nicefrac{16}{3}$. Results are also shown where the simulation domain
  is extended by a factor of 1.5 in either or both dimensions. In the extended
  simulations, the number of grid points is increased to keep the grid spacings
  $\Delta x=\Delta y$ the same.\label{fig:drop-domain_size}}
\end{figure}

\section{Tests of the simulation accuracy}
\label{app:drop-accur}
The simulation domain size, which is set via the non-dimensional parameter
$\tilde{L}$ in \autoref{tbl:drop-sim_params} and the domain aspect ratio $\beta$,
may affect on the results. Since the boundary conditions on the top and
right boundaries are based on an inviscid assumption, the domain size
affects the extent to which viscosity is resolved. In addition, when solving
for the pressure in the gas layer, the boundary condition of $p=P_0$ is imposed
at $x=L$, and thus extending the domain affects the influence of this boundary
condition on the simulation.

To test the sensitivity of the simulation results to the domain size, we
performed simulations where either the horizontal dimension, vertical
dimension, or both dimensions were extended by a factor of 1.5. In each of
these simulations the grid spacings $\Delta x = \Delta y$ were kept the same,
so that the horizontal extension increases the grid points from 5120 to 7680,
and the vertical extension increases the grid points from 960 to 1440.
\autoref{fig:drop-domain_size} shows the height profiles in the thin gas layer for
the four simulations, for (a) $\nu_l=\vten$ and (b)
$\nu_l=\vhun$. The vertical extensions give visually
indistinguishable curves, indicating that vertical dimension is sufficiently
large to resolve the viscous effects even for the larger value of $\nu_l$. The
horizontal extensions create a minor vertical shifts in the curves, suggesting
that the pressure boundary condition has an effect on the results. However,
these shifts are still acceptably small, and result in no substantial change to
the position and time at which lift-off occurs.

\begin{figure}
  \centering
  \includegraphics[width=\linewidth]{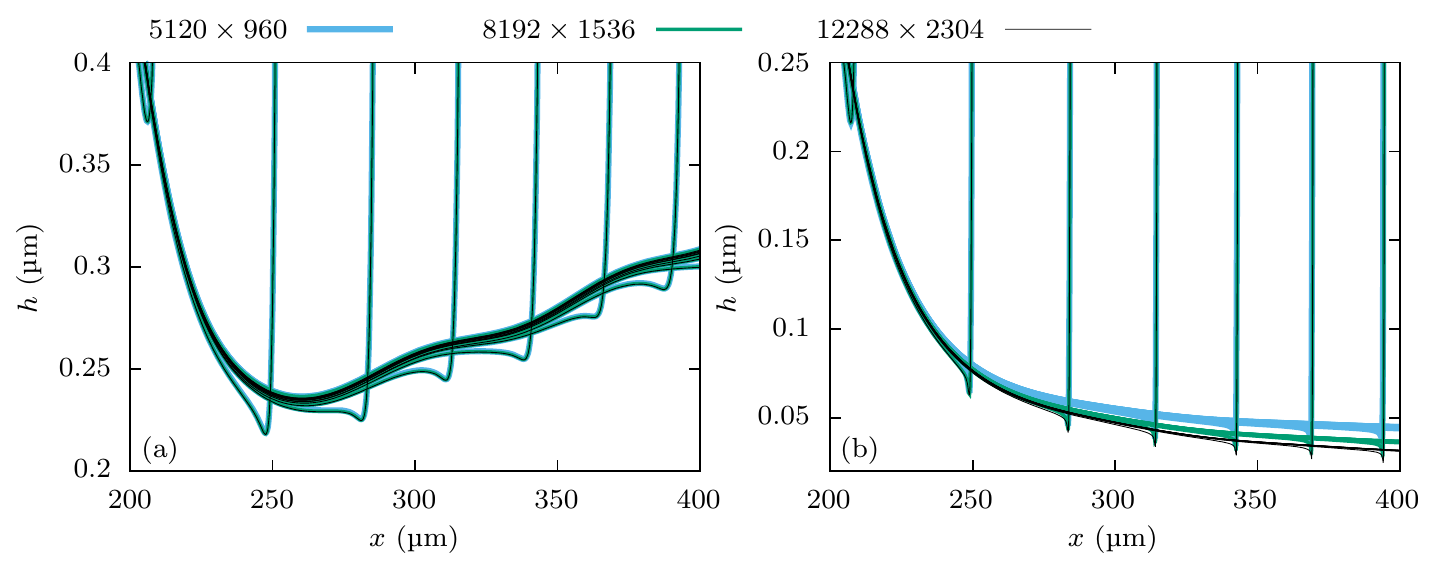}
  \caption{Plots of the height profiles spaced $6.004\,\text{\textmu{}s}$ apart
  for three different resolutions using the baseline parameters and liquid
  viscosity $\nu_l =\vten$, with (a) surface tension $\sigma
  =0.072\,\text{N}/\text{m}$, and (b) zero surface tension.\label{fig:drop-conv_sig}}
\end{figure}

We also examined the sensitivity of the results to the numerical grid size.
\autoref{fig:drop-conv_sig}(a) shows height profiles for a simulation using the
baseline parameters with liquid viscosity $\nu_l=\vten$ and surface
tension $\sigma=0.072\,\text{N}/\text{m}$. Simulations with higher
resolutions of $8192\times1536$ and $12288\times 2304$ are also plotted,
resulting in visually indistinguishable results. \autoref{fig:drop-conv_sig}(b)
shows height profiles when the surface tension is set to zero and all other
parameters are kept the same. In this case, as discussed in
\autoref{sec:drop-surf_ten}, the leading tip becomes very sharp since the
regularizing effect of surface tension is removed. As the tip moves across the
grid, there will be a resolution-dependent numerical diffusion that will act as
an effective small surface tension. Thus in this case there is a small shift in
the height profiles. While this remains small, it makes the precise behavior of
the leading tip difficult to resolve for small values of $\nu_l$, requiring
larger grid sizes as presented in \autoref{sec:drop-surf_ten}.

\section{Additional model fits for the lift-off time}
\label{sec:drop-app-model-fits}
\begin{figure}
  \begin{center}
    \includegraphics[width=\textwidth]{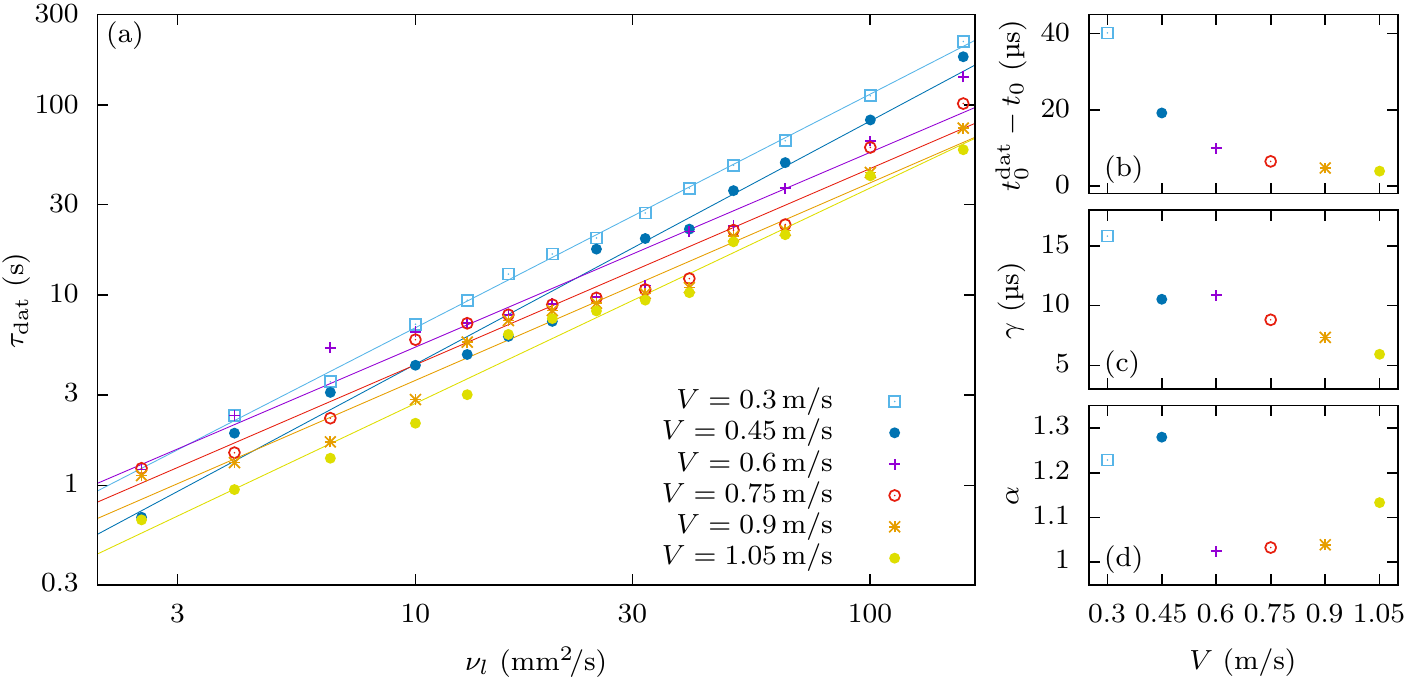}
  \end{center}
  \caption{(a) Model fits for minimizing the residual $S$ for the three
  parameters $(t_0^\text{dat},\gamma,\alpha)$. The straight lines show the
  fitted model $\taudat = \gamma (\nu_l/\nu_\text{sc})^\alpha$. (b--d) values of
  the parameters $\tzdat$, $\gamma$, and $\alpha$.\label{fig:supp_dat1}}
\end{figure}

In \autoref{sec:drop-visc_liftoff} of the paper, we examine how the lift-off time $t$ varies as a
function of liquid viscosity $\nu_l$. For a sequence of measurements
$(\nu_{l,k},t_k)$ we fit the model $\taudat = t -\tzdat = \gamma
(\nu_l/\nu_\text{sc})^\alpha$ for the three parameters
$(\tzdat,\gamma,\alpha)$. Here $\nu_\text{sc}=\vtwenty$ is an arbitrary
viscosity scale. We minimized the residual
\begin{equation}
  \label{eqn:supp_resid}
  S(\tzdat,\gamma,\alpha) =\frac{1}{2} \sum_k \left( \log \gamma + \alpha \log \frac{\nu_{l,k}}{\nu_\text{sc}} - \log(t_k-\tzdat)\right)^2,
\end{equation}
using the L-BFGS-B algorithm for bound-constrained nonlinear
optimization~\cite{byrd95}, and we enforced $t_0^\text{dat}<0.999 t_\text{min}$
and $\alpha>0$, where $t_\text{min} = \min_k \{t_k\}$. \autoref{fig:supp_dat1}
shows the results of this optimization for the six different initial drop
velocities $V$, along with the values of the fitted parameters. The mean
value of the residual is $\bar{S}=0.254$.

\begin{figure}
  \begin{center}
    \includegraphics[width=\textwidth]{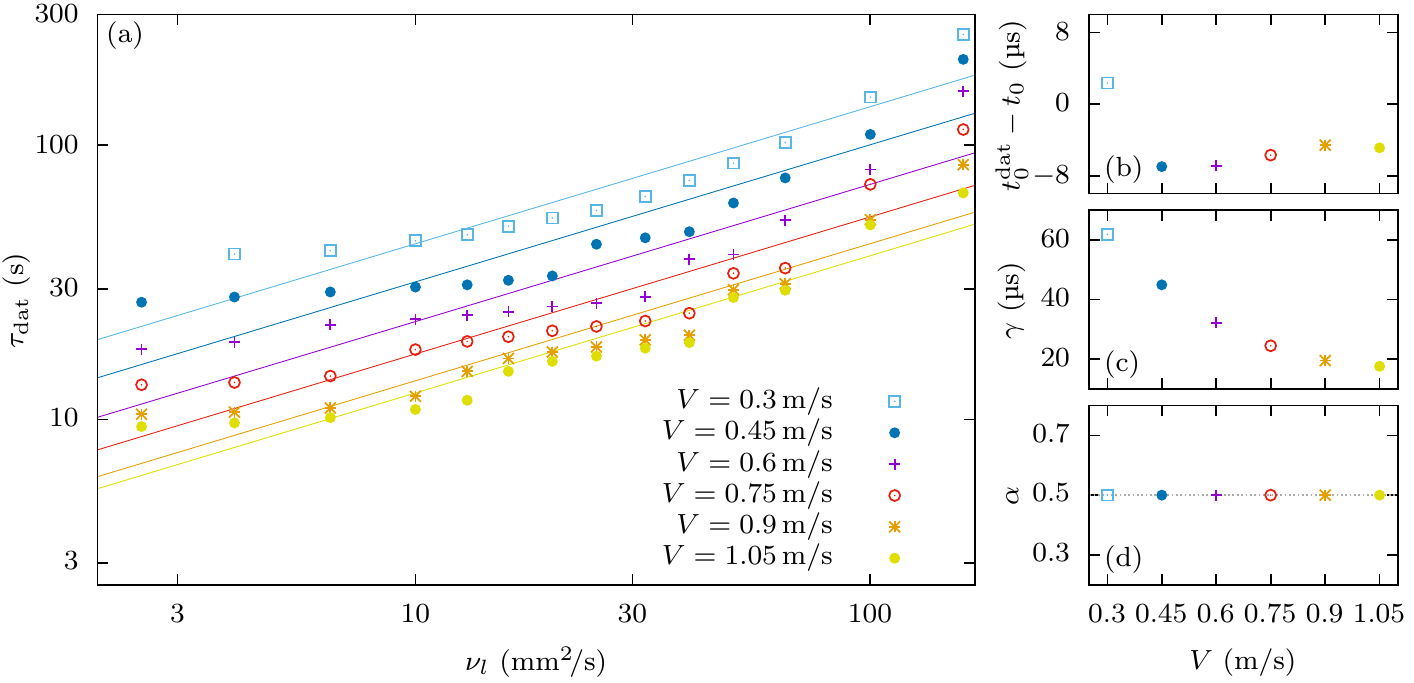}
  \end{center}
  \caption{(a) Model fits for minimizing the residual $S$ for the two
  parameters $(t_0^\text{dat},\gamma)$ where $\alpha=\nicefrac12$. The straight
  lines show the fitted model $\taudat = \gamma (\nu_l/\nu_\text{sc})^\alpha$.
  (b--d) values of the parameters $\tzdat$, $\gamma$, and $\alpha$, with the
  dotted line indicating the constrained parameter.\label{fig:supp_dat2}}
\end{figure}

\begin{figure}
  \begin{center}
    \includegraphics[width=\textwidth]{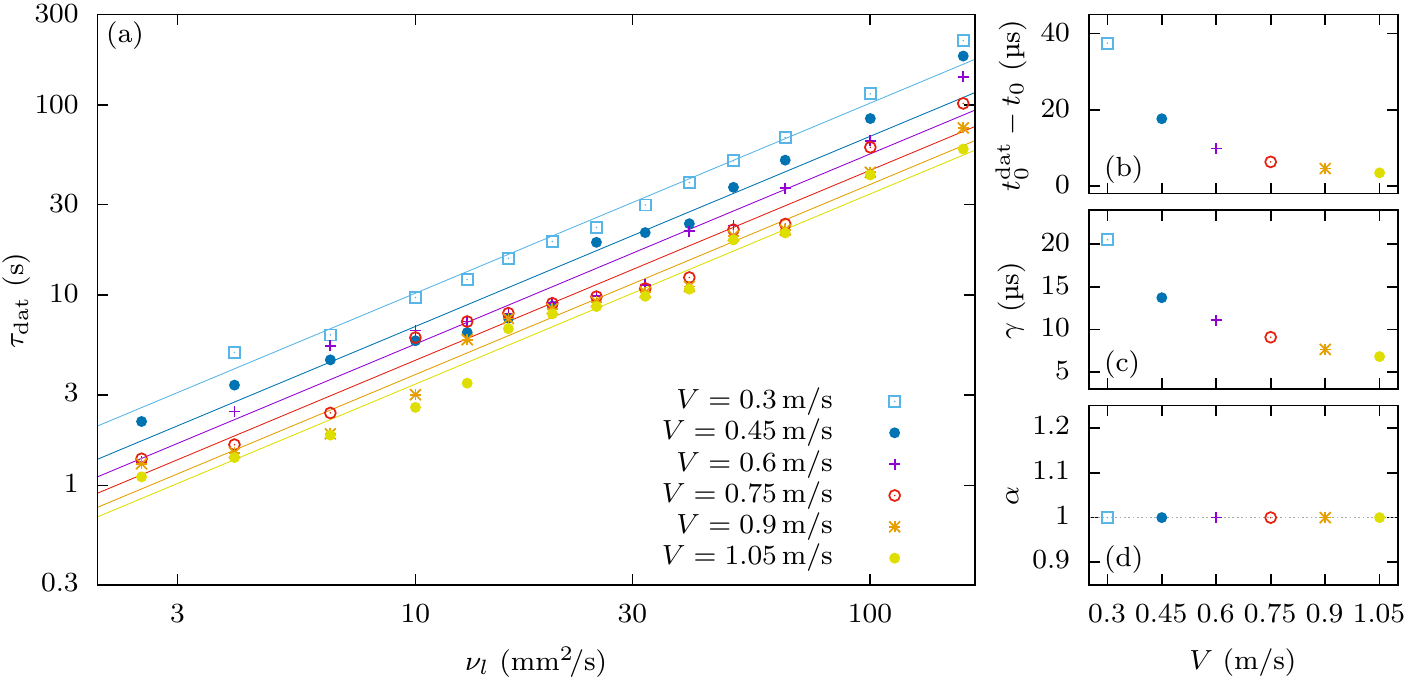}
  \end{center}
  \caption{(a) Model fits for minimizing the residual $S$ for the two
  parameters $(t_0^\text{dat},\gamma)$ where $\alpha=1$. The straight lines
  show the fitted model $\taudat = \gamma (\nu_l/\nu_\text{sc})^\alpha$. (b--d)
  values of the parameters $\tzdat$, $\gamma$, and $\alpha$, with the dotted
  line indicating the constrained parameter.\label{fig:supp_dat3}}
\end{figure}

\begin{figure}
  \begin{center}
    \includegraphics[width=\textwidth]{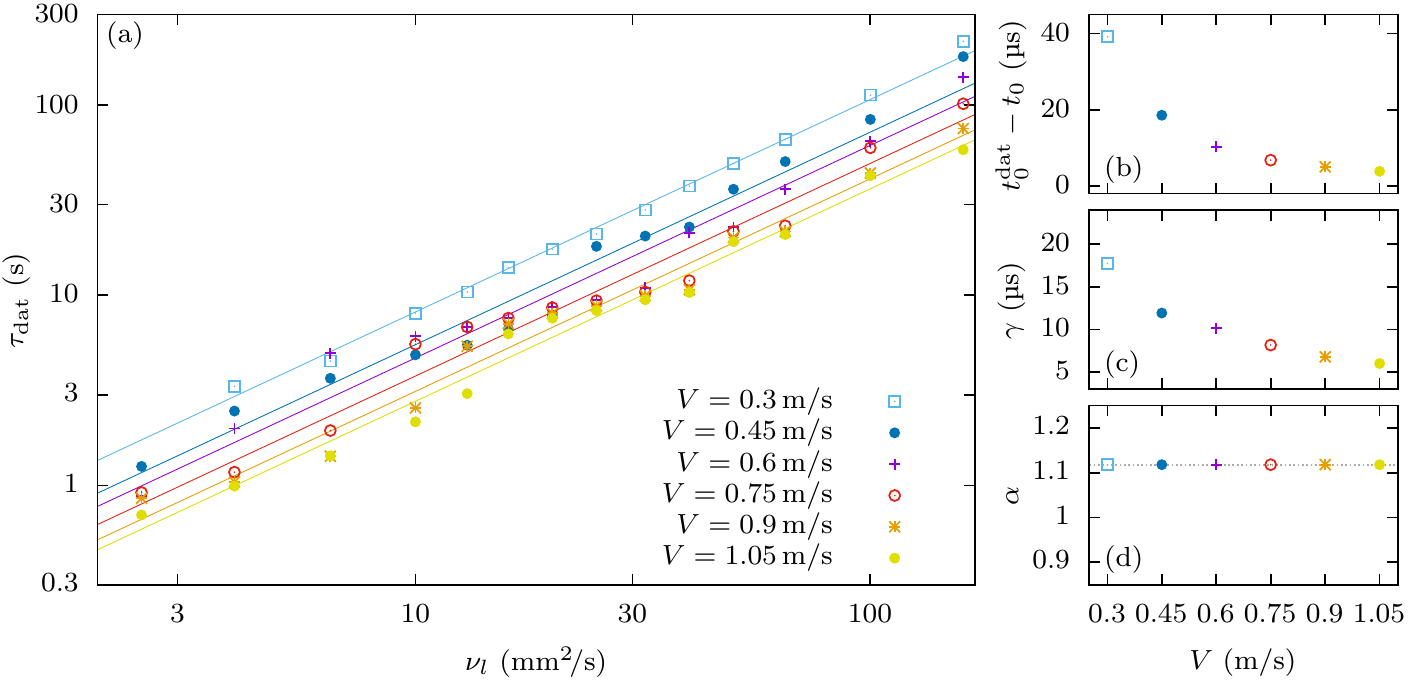}
  \end{center}
  \caption{(a) Model fits for minimizing the residual $S$ for the two
  parameters $(t_0^\text{dat},\gamma)$ where $\alpha=1.1181$. The straight
  lines show the fitted model $\taudat = \gamma (\nu_l/\nu_\text{sc})^\alpha$.
  (b--d) values of the parameters $\tzdat$, $\gamma$, and $\alpha$, with the
  dotted line indicating the constrained parameter.\label{fig:supp_dat4}}
\end{figure}

In the experiments by \cite{kolinski2014lift}, the
parameter $\alpha$ is proposed to be $\nicefrac12$. We therefore ran an
additional optimization where $\nu_l$ is constrained to be $\nicefrac12$. The
mean residual increases to $\bar{S}=0.385$ and the model does not capture
the curve in the data points (\autoref{fig:supp_dat2}). We ran a further
optimization where $\alpha=1$ (\autoref{fig:supp_dat3}), which results in
$\bar{S}=0.305$. We found that the value of $\alpha$ that minimizes the mean
residual is $\alpha=1.1181$. For this case $\bar{S}=0.281$ and the optimization
results shown in \autoref{fig:supp_dat4}.

\begin{figure}
  \begin{center}
    \includegraphics[width=\textwidth]{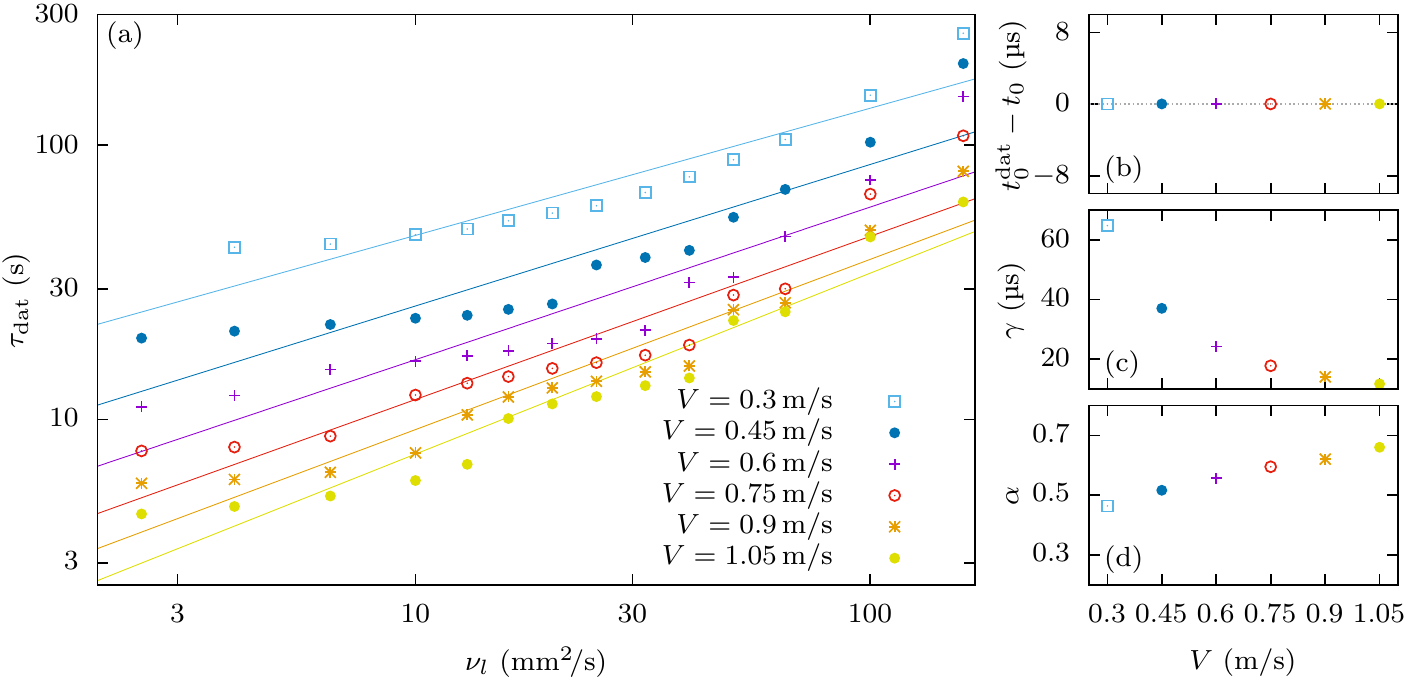}
  \end{center}
  \caption{(a) Model fits for minimizing the residual $S$ for the two
  parameters $(\gamma,\alpha)$ where $t_0^\text{dat}=t_0$. The straight
  lines show the fitted model $\taudat = \gamma (\nu_l/\nu_\text{sc})^\alpha$.
  (b--d) values of the parameters $\tzdat$, $\gamma$, and $\alpha$, with the
  dotted line indicating the constrained parameter.\label{fig:supp_dat5}}
\end{figure}

\begin{figure}
  \begin{center}
    \includegraphics[width=\textwidth]{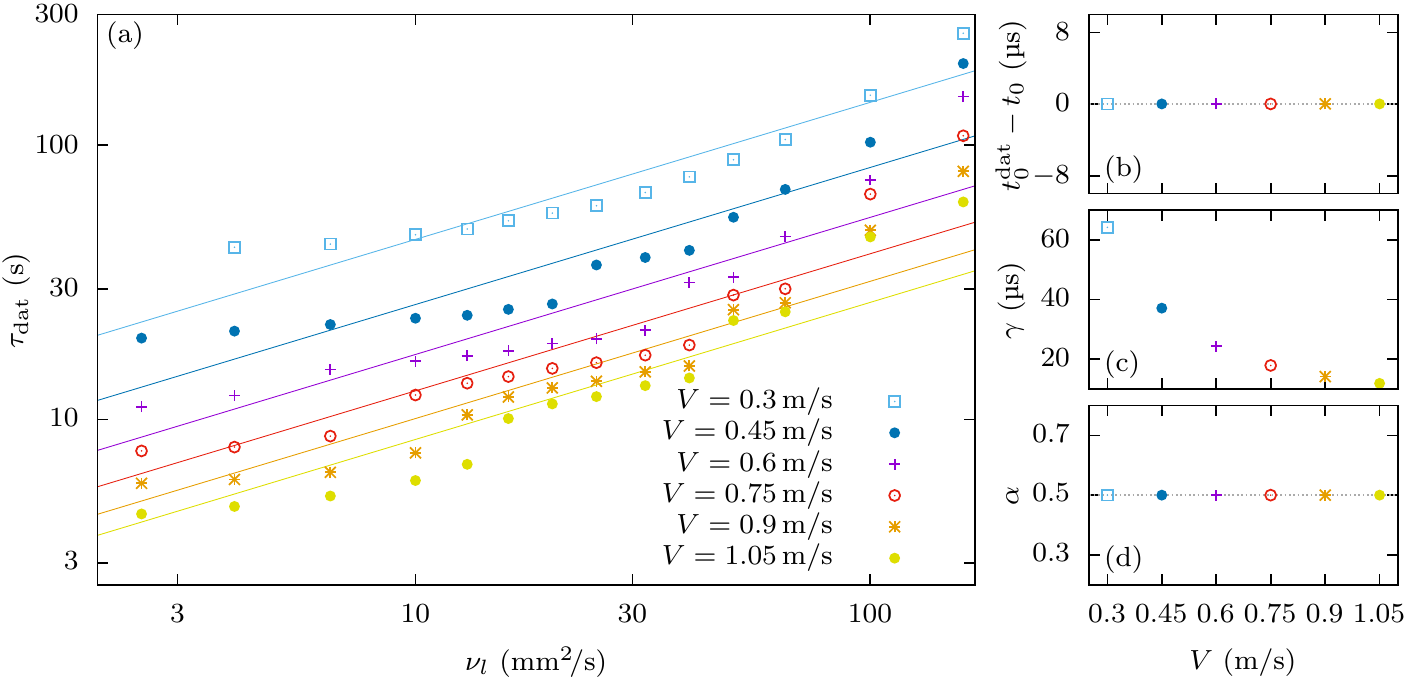}
  \end{center}
  \caption{(a) Model fits for minimizing the residual $S$ for the 
  paramater $\gamma$, where $t_0^\text{dat}=t_0$ and $\alpha=\nicefrac12$. The
  straight lines show the fitted model $\taudat = \gamma
  (\nu_l/\nu_\text{sc})^\alpha$. (b--d) values of the parameters $\tzdat$,
  $\gamma$, and $\alpha$, with the dotted lines indicating the constrained
  parameters.\label{fig:supp_dat6}}
\end{figure}

\autoref{fig:supp_dat5} shows an optimization when the time origin is fixed to
the original proposed definition of $t_0=H/V$, where $H$ is the drop height
above the surface. In \autoref{fig:supp_dat6} the exponent is additionally
constrained to $\alpha=\nicefrac12$. For each of the six optimizations, the
values of $S$ are reported in \autoref{tab:svals}.

\begin{table}[b]
  \begin{center}
    \small
    \begin{tabular}{l|llllll|l}
      Constraints & \Sc{0.3} & \Sc{0.45} & \Sc{0.6} & \Sc{0.75} & \Sc{0.9} & \Sc{1.05} & $\bar{S}$ \\
      \hline
      -- & 0.019 & 0.260 & 0.415 & 0.356 & 0.264 & 0.207 & 0.2536 \\
      $\alpha=\nicefrac12$ & 0.226 & 0.483 & 0.523 & 0.461 & 0.332 & 0.287 & 0.3855 \\
      $\alpha=1$ & 0.093 & 0.466 & 0.417 & 0.358 & 0.266 & 0.229 & 0.3049 \\
      $\alpha=1.1181$ & 0.039 & 0.345 & 0.446 & 0.372 & 0.273 & 0.208 & 0.2805 \\
      $\tzdat = t_0$ & 0.218 & 0.520 & 0.572 & 0.494 & 0.351 & 0.314 & 0.4116 \\
      $\tzdat = t_0, \alpha=\nicefrac12$ & 0.227 & 0.522 & 0.602 & 0.578 & 0.488 & 0.553 & 0.4952
    \end{tabular}
  \end{center}
  \caption{Values of the residual from equation \eqref{eqn:supp_resid} for the
  six different optimizations considered. The first column shows the additional
  constraints that are applied on the three parameters
  $(t_0^\text{dat},\gamma,\alpha)$. The next six columns show the residual for
  the six different initial drop velocities $V$. The final column shows the
  mean residual taken over the six values of $V$.\label{tab:svals}}
\end{table}

\clearpage
\bibliography{library}
\end{document}